%% file: main.tex
\pdfoutput=1

\documentclass[a4paper,12pt, oneside]{book}
\input{chapters/cover/packages.tex}

\makeatletter
\renewcommand{\@dotsep}{10000} 
\makeatother

\usepackage{graphicx}
\usepackage[firstpage=true]{background}
\backgroundsetup{scale=0.5, angle=35, opacity=0.04, position={27cm, -42cm}, 
contents={\includegraphics[]{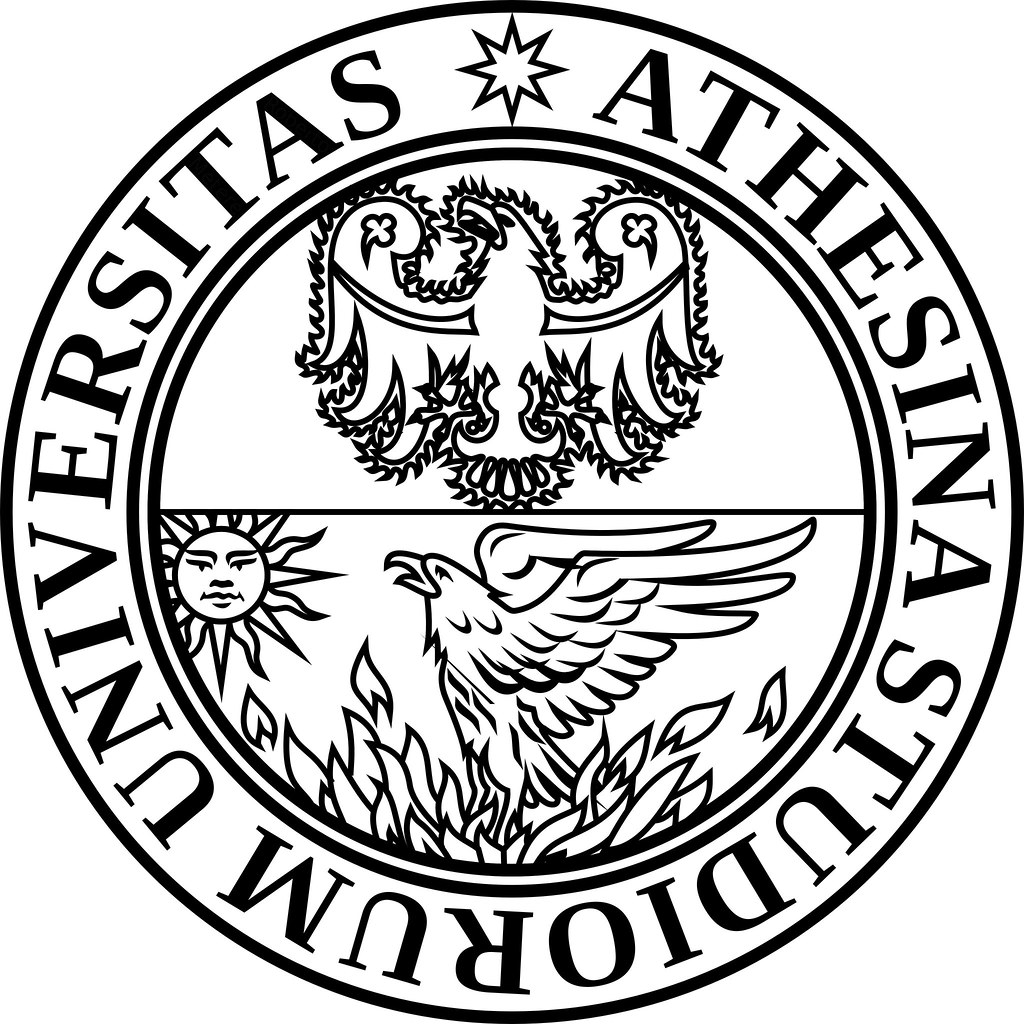}}}

\begin{document}

\begin{spacing}{1.15}
\input{chapters/cover/cover.tex}
\pagenumbering{arabic}
\end{spacing}

\begin{spacing}{1.25}

\chapter{Abstract}
\input{chapters/Abstract/abstract.tex}

\chapter{Objectives}
\input{chapters/Objectives/objectives.tex}

\chapter{Introduction}
\input{chapters/introduction/introduction.tex}

\chapter{Results}
\input{chapters/Results/results.tex}

\chapter{Methods and Resources}
\input{chapters/Methods/methods.tex}

\chapter{Discussion and Conclusions}
\input{chapters/Conclusions/conclusions.tex}

\chapter{Data availability}
\input{chapters/Data/data.tex}

\twocolumn
\renewcommand*{\bibname}{References}
\printbibliography

\onecolumn 
\chapter{Supplementary figures}
\input{chapters/Supplementary/supplementary.tex}

\end{spacing}
\end{document}

%% file: chapters/cover/packages.tex
\usepackage[utf8]{inputenc}
\usepackage{graphicx}
\usepackage[paper=a4paper, margin={20mm, 25mm}]{geometry}
\usepackage[sorting=none]{biblatex}
\usepackage{color}
\usepackage[colorlinks = true,
            linkcolor = blue,
            urlcolor  = blue,
            citecolor = blue,
            anchorcolor = blue]{hyperref}
\usepackage[section]{placeins}
\usepackage[font=small]{caption}
\counterwithout{figure}{chapter}

\usepackage{subcaption}
\usepackage{graphicx}
\DeclareCaptionFormat{custom}
{%
    \textit{\textbf{#1#2}}#3
}
\usepackage{titlesec}
\titleformat{\chapter}[display]{\fontsize{16}{30}\bfseries}{}{}{}

\usepackage{sectsty}
\chapterfont{\fontsize{15}{15}\selectfont}
\sectionfont{\fontsize{14}{20}\selectfont}
\subsectionfont{\fontsize{13}{15}\selectfont}
\subsubsectionfont{\fontsize{12}{10}\selectfont}

\usepackage{fancyhdr} 
\titleformat{\chapter}[display]
  {\normalfont\bfseries}{}{10pt}{\huge}

\titleformat{\chapter}%
  {\normalfont\bfseries\Huge}{\thechapter.}{10pt}{}

\usepackage[ruled,linesnumbered]{algorithm2e}
\usepackage{amsmath}
\usepackage{cleveref}

\usepackage{amsfonts}

\usepackage{setspace}

\usepackage{xcolor}
\newcommand\rurl[1]{%
  \href{http://#1}{\nolinkurl{#1}}%
}

%% file: chapters/cover/cover.tex
\pagenumbering{gobble}
\vspace{3em}
\begin{figure}[!ht]
    \centering
    \includegraphics[width=0.25\textwidth]{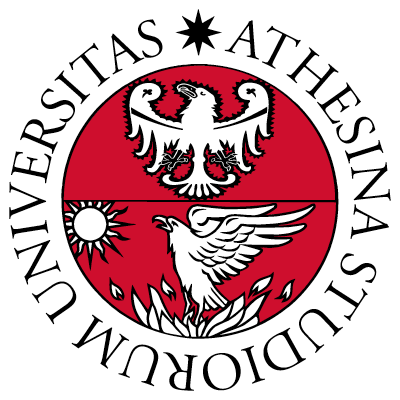}
\end{figure}

\vspace{2em}
\begin{center}
    \Large\textsc{University of Trento}
\end{center}

\begin{center}
    \large\textsc{Master of Science in Quantitative and Computational Biology}
\end{center}

\vspace{1em}
\begin{center}
    Academic year 2022--2023
\end{center}

\vspace{4em}

\begin{center}

\hrule
    \vspace{2em}
    {\Large\textbf{\textsc{Discovery and optimization of cell-type-specific DNA methylation markers for \textit{in silico} deconvolution}}}
    \vspace{2em}
\hrule
\end{center}

\vspace{4em}

\hspace{-1.5em}\textbf{Author:}
\hfill
\textbf{Supervisor:} \\
Aleksa Krsmanović
\hfill
Francesca Demichelis \\

\hfill \textbf{Co-supervisor:} \\
\hspace*{0pt}\hfill Gian Marco Franceschini

\vspace{4em}
\begin{center}
    \textsc{CIBIO - Center for Integrative Biology}
\end{center}

\begin{center}
    \textsc{Laboratory of Computational and Functional Oncology}
\end{center}

\vspace{2em}
\begin{center}
    22nd March, 2023
\end{center}

\makeatletter
\renewcommand{\@dotsep}{1} 
\makeatother
\hspace{0em}
\vspace{0em}
{
  \hypersetup{linkcolor=black}
  \tableofcontents
}
\newpage

%% file: chapters/Abstract/abstract.tex


The tissue-specific nature of epigenetic control is the greatest driver of cell-type heterogeneity. DNA methylation is no exception and has been implicated in various regulatory processes ranging from cell differentiation to imprinting. As the methyl group is embedded in the DNA molecule, assessing DNA methylation is particularly promising in liquid-biopsy-based approaches, as cell-free DNA retains information related to its cell of origin. In this work, I leverage a recently profiled collection of cell-sorted whole genome bisulfite profiles of 44 healthy cell types. The high quality and purity of such data provide an ideal basis for discovering and characterizing discriminative DNA methylation regions that could serve as a reference for \textit{in silico} deconvolution. First, I characterize differentially methylated regions between every pair of cell types, obtaining a meaningful measure of divergence. Pairwise differences were then aggregated to identify a set of uniquely (de)methylated regions (UMRs) for each cell type. Identified UMRs are predominantly hypomethylated and their numbers vary greatly across cell types. They are mostly located in enhancer regions and strongly support cell-type-specific characteristics. As mapping onto UMRs has proven unsuitable for deconvolution, I developed a novel approach utilizing the set cover algorithm to select discriminative regions for this purpose. Based on these regions, deconvolution was performed in two distinct approaches: a beta-value-based and a read-level one. Both approaches outperform an existing deconvolution software modeled on the same data 3-fold in terms of total deconvolution error. Surprisingly, the beta-based approach slightly outperformed the read-level one. Overall, I present an adaptable, end-to-end software solution (methylcover) for obtaining accurate cell type deconvolution, with possible future applications to non-invasive assays for disease detection and monitoring.

%% file: chapters/Objectives/objectives.tex
 The interpretation of DNA methylation data requires a clear understanding of cell-type heterogeneity. 
 This is particularly relevant in the context of cell-type deconvolution, the project's primary task aimed at resolving the relative abundances of single cell types accounting for an observed admixed signal.
 Motivated by this problem, here I describe an end-to-end workflow that aims to improve upon current feature selection strategies previously applied in this context. In particular, the thesis is divided into two main sections with distinct aims:
 
\noindent 1. Identify unique DNA methylation markers that uniquely characterize healthy cell types of interest

\noindent 2. Optimize the set of markers to provide robust deconvolution results from DNA methylation profiles of bulk samples

%% file: chapters/introduction/introduction.tex
\section{The human epigenome}


Despite the fact that every cell in the human body carries identical genetic information, their phenotypes vary due to differences in both the qualitative and quantitative aspects of their gene expression landscapes. These expression patterns are formed during development and perserved as cells divide via mitosis, uniquely characterizing different cell types.
Epigenetic regulation, such as 1) post-transnational histone modification, 2) DNA methylation, and 3) RNA-dependent mechanisms, are known to stably alter gene expression by acting on the transcriptional level \cite{nature_intro_epi}. This epigenetic information, which is passed down to daughter cells during mitosis and distinguishes different cell types, is collectively termed the epigenome. 
While an individual's genome is generally considered to be stable throughout their lifespan and largely the same in all cells, the epigenome exhibits variation between different cells types and is subject to alterations over time, influenced by both internal and external factors \cite{nature_intro_epi}.
The epigenetic landscape of various cell types is prone to vast modifications during certain biological stages like fetal development, puberty, or pregnancy. Importantly, its changes can also be indicative, or even the cause of a number of diseases and conditions \cite{esteller2008epigenetics}. For this reason, understanding epigenetic mechanisms has been a main focus of studies concerning healthy aging, developmental biology, and oncology in the last decade.

\section{DNA methylation}
DNA methylation is an epigenetic mechanism that refers to the reversible addition of a methyl group (CH\textsubscript{3}) on the 5' position of cytosine residues giving rise to 5-methylcytosine (5-mC). Cytosine methylation can be mono or biallelic and is inherited by the daughter cell after mitosis. In mammals, this modification almost exclusively happens when cytosine is located 5' to guanine -- denoted as CpG methylation (p stands for the phosphodiester bond linking the two nucleotides). Less frequently, DNA methylation can also occur in non-CpG contexts -- CHH and CHG (where H substitutes one nucleotide: either adenine, cytosine, or thymine) \cite{jang2017cpg}. These non-CpG modifications are quite rare in human somatic cells (at around 0.02\%). However, in pluripotent cells, they can amount to up to 25\% of the total DNA methylation \cite{laurent2010dynamic}. The addition of methyl groups changes the biophysical properties of DNA by inhibiting interactions with specific proteins while facilitating the binding of others. However, in general, it is associated with reduced gene expression when present in functional elements such as promoters or enhancers.

Places in the genome where CpG pairing occurs at a higher frequency are called CpG islands. These regions are frequently located in gene promoters, but recent studies have shown that about 50\% of them are located in inter or intragenic regions \cite{illingworth2008novel}. 
Although in healthy cells, 70--80\% of CpG sites are methylated, the majority of CpG islands located at promoter sites have low levels of methylation, thus allowing gene expression. Methylated CpG islands at promoters are associated with long-term silencing such as imprinted genes and X chromosome inactivation \cite{strichman2002genome, cross1995cpg}.

CpG shores are regions stretching 2kb upstream and downstream of CpG islands. CpGs in CpG shores are often less dense and more methylated w.r.t. CpG islands. One previous investigation demonstrated that gene expression levels were negatively associated with methylation levels at CpG island shores \cite{irizarry2009human}. Beyond CpG shores, open-sea CpG sites are even more sparse and most often intergenic and fully methylated.


DNA methylation can affect gene expression via three main mechanisms. Firstly, it can change the chromatin structure and accessibility of transcription factors; secondly, by changing the binding affinity of transcription factors to binding sites at gene promoters and enhancers and lastly, by affecting the binding affinity of methylation-specific recognition proteins. For example, methyl-binding MeCP proteins, which associate with various transcriptional repressors, bind methylated CpG sites where they can further reduce transcription by condensing the chromatin by recruiting histone deacetylases \cite{lee2020key}.

\subsection{DNA methylation establishment and maintenance}

The creation and maintenance of DNA methylation are performed by a family of enzymes called DNA methyltransferases (DNMTs). In mammals, five members of this family have been identified: DNMT1, DNMT2, DNMT3a, DNMT3b, and DNMT3L \cite{weber2007genomic}. DNMT1 plays a crucial role in preserving DNA methylation during mitosis by replicating the methylation patterns of the parent DNA strand onto the newly synthesized one. DNMT3a and DNMT3b are responsible for \textit{de novo} DNA methylation, as well as assisting DNMT1 to propagate methylation patterns during cell division \cite{kulis2010dna}. DNMT3L has been shown to stimulate DNMT3a activity and reduce gene expression by interacting with histone deacetylase 1 \cite{chedin2002dna}. In \textit{in vitro} experiments, DNMT2 has been shown to have weak DNA methylation ability, but higher RNA methylation capability \cite{schaefer2010solving}.

DNA demethylation may occur either passively when methylation patterns fail to be maintained after cell division, or actively when methyl groups are cleaved off. Passive DNA demethylation refers to the loss of methyl groups when DNMT1 is inhibited or absent during DNA replication \cite{wu2010active}. Active mammalian DNA demethylation remains unclear as several mechanisms seem to be complementing each other. For example, active demethylation can be observed in zygotes before the first meiotic division when both paternal and maternal genomes are rapidly demethylated \cite{guo2014active}. The family of ten-eleven translocation (TETs) enzymes and thymine DNA glycosylase are key factors in both active and passive demethylation \cite{onodera2021roles}.

\subsection{DNA methylation and diseases}

Proper establishment and maintenance of DNA methylation are essential for normal development in mammals. Studies have shown that knocking-out DNMT1 or DNMT3b in mice results in embryonic lethality and that mice lacking DNMT3a die within a few weeks of birth \cite{okano1999dna, bird2002dna}.

A well-studied category of conditions induced by aberrant DNA methylation is imprinting disorders. Genomic imprinting refers to a parent-of-origin-specific epigenetic modification where gene expression of the imprinted allele becomes silenced while only the other allele is expressed. Examples of these disorders include developmental syndromes such Beckwith-Wiedemann syndrome, Prader-Willi syndrome, and transient neonatal diabetes mellitus\cite{debaun2003association,goldstone2004prader,temple2002transient}. Moreover, a wide range of diseases have been reported to be associated with defective methylation patterns including rheumatoid arthritis \cite{liu2013epigenome}, autism \cite{ladd2014common}, and Alzheimer’s disease \cite{de2014alzheimer}.

In cancer, distorted DNA methylation patterns
both accompany genic oncological mechanisms and it is even hypothesized that they can induce tumorigenesis \cite{yu2014targeted}. Global hypomethylation of DNA is a hallmark of malignancies and can be frequently observed in many cancers. This loss of methylation usually occurs in repetitive sequences, transposable elements, and regulatory regions and leads to genomic instability and chromosomal abnormalities \cite{esteller2007epigenetic}. On the other hand, hypermethylation of CpG islands in promoter regions of tumor suppressor genes can lead to their silencing, thus allowing cancer cells to proliferate unchecked \cite{park2011promoter}. Overall, vast changes in the methylome are a common feature of cancer cells and might play an important role in cancer development and progression. Therefore, DNA methylation has emerged as a promising target for cancer diagnosis, prognosis and treatment, with several drugs already developed that target DNA methylation enzymes to restore normal gene expression and inhibit tumor growth \cite{issa2007dna}.

\subsection{Plasma cell-free DNA methylation}

Plasma cell-free DNA (cfDNA) is the DNA that circulates freely in the bloodstream (i.e. it is not included in cells or other compartments), originating from different tissues. As the methyl group is embedded into the DNA molecule, it is possible to trace the origin of captured fragments using its distribution patterns \cite{diaz2014liquid}. 

Recently, cfDNA has emerged as a promising biomarker for non-invasive early detection and monitoring of various diseases in which tissue degradation occurs, including cancer. Obtaining such markers is usually performed by identifying methylation patterns able to discriminate the condition of interest \textit{in situ}, then capturing those regions in blood using high-throughput sequencing or other methods. Apart from early detection, cfDNA methylation has the potential to be used as a biomarker for cancer prognosis and treatment response \cite{chimonidou2013cst6}. By analyzing the changes in cfDNA methylation over time, physicians can assess the effectiveness of cancer treatments and adjust the treatment plan as needed. Moreover, cfDNA methylation analysis has shown promise in detecting minimal residual disease in cancer patients \cite{peng2021circulating}. The possibility may allow for earlier and more effective interventions to prevent disease recurrence. 
In the non-cancer context, cfDNA methylation has shown promise as a prognostic tool for organ transplant patients \cite{boer2021variations}. By analyzing the abundance of the grafted organ DNA in blood over time, physicians can assess the risk of organ rejection and adjust the treatment plan accordingly.
Overall, the use of cfDNA methylation is rapidly evolving, the cost of profiling is decreasing, and it is expected to become a commonly used tool in clinical practice for detecting, diagnosing, and monitoring various conditions in the near future.

\section{Measuring and analyzing DNA methylation}

\subsection{Common assays}

Historically, a number of assays have been used to detect methylation. Depending on the investigation objective and resources available, a researcher will most likely choose one of the following:
\begin{enumerate}
  \item Bisulfite microarrays
  \item Bisulfite next-generation sequencing 
  \item Immuno-precipitation-based sequencing (MeDiP-seq)
\end{enumerate}
\vspace{0.5em}
Both next-gen sequencing methods and array-based ones require a step of bisulfite conversion. The bisulfite conversion of DNA changes unmethylated cytosines into uracils through deamination while leaving 5-mC and 5-hydroxymethylcytosine (5-hmC -- the oxidation product of the former) unchanged. The DNA treated with bisulfite undergoes amplification using PCR, leading to the conversion of uracils to thymines. Consequently, unmethylated cytosines in the DNA molecule are transformed into thymines while the methylated cytosines remain as cytosines. This process results in a distinct nucleotide variation between methylated and unmethylated cytosines that can be conveniently detected when compared to the initial reference genome.
A disadvantage of all assays involving bisulfite treatment is that it becomes impossible to differentiate the methylated cytosine and its oxidized product. However, 5-hmC distribution across different human cell types is low w.r.t 5-mC and ranges from 0.04 -- 0.6\% \cite{li2011distribution}. 

Bisulfite microarrays aim to identify the ratio of methylated vs unmethylated CpGs at predetermined genomic locations. Most widely used panels capture around 27000, 450000, and 850000 such CpGs (Illumina's Infinum HumanMethylation27, Infinum HumanMethylation450 and Infinum HumanMethylation850 BeadChip arrays, respectively). Even with the HumanMethylation450 array, 99\% RefSeq genes (hg19 reference) and 96\% CpG islands are covered \cite{sun2015base}. Here, two types of fluorescent antibodies (each with a specific wavelength) are added to the panel probes where they bind the converted or unconverted cytosine sequences with high selectivity. Once the ratio of luminescence of these two antibodies is determined, it can be used as a proxy for the methylation level of a specific site. This value is called a beta value, and it corresponds to the ratio of methylated cytosines at a CpG site. A disadvantage of microarray assays is the fact that they are limited to a small number of predetermined CpGs out of $\sim$28.2 million in the human genome. However, they are quite affordable, provide reliable beta values, and, if needed, a custom panel covering CpGs of choice can be crafted.  

Bisulfite next-gen sequencing, on the other hand, allows greater breadth and single-nucleotide resolution DNA methylation. One can sequence the whole genome, identifying the methylation status of virtually all CpGs -- termed whole genome bisulfite sequencing (WGBS). In this assay, the reliability of inferred methylation values is proportional to the genomic coverage. However, the experiment cost also scales with it. Depending on financial resources, one might choose 1) an expensive high-coverage genome-wide experiment or 2) a cheaper genome-wide low-pass one. If only specific genomic regions are of interest, 3) targeted sequencing can provide high coverage data while also being affordable by focusing the assay on only the desired regions. A commonly used targeted assay for standard, pre-selected genomic regions is called 4) reduced representation bisulfite sequencing (RRBS). Here, by using specific restriction enzyme digestions and size selection, one can get single-nucleotide resolution DNA methylation data for $\sim$80\% of human CpG islands and more than 60\% of human promoters while only sequencing $\sim$3\% of the genome \cite{sun2015base}.

In sequencing experiments, the aggregation of adjacent CpG site methylation values is usually done to obtain a mean methylation estimate of a region. This provides a more stable estimate than observing a single CpG, which might not reflect methylation in neighboring sites.

MeDIP-seq is a genome-wide assay that combines immunoprecipitation of methylated DNA fragments with next-generation sequencing to generate genome-wide DNA methylation maps. This method is based on the use of an antibody specific to 5-mC, but not 5-hmC. After immunoprecipitation, the methylated DNA fragments are purified and subjected to next-generation sequencing. MeDIP-seq has several advantages over bisulfite sequencing, including high sensitivity, selectivity for 5-mC only, and lower cost. However, MeDIP-seq infers methylation in lower resolution (block of around 100bp) compared to WGBS, which can provide single-base resolution \cite{rauch2009human}.

For this research project, only bisulfite sequencing assays are of interest. 

\subsection{Identifying differential and unique methylation}

Upon obtaining single CpG site methylation values for samples in test and control groups, the goal is often to identify which sites or regions are deferentially methylated. Identifying deferentially methylated sites (DMSs) is a relatively easy task, which can be solved with a number of statistical tests or simply by thresholding the difference of intra-group methylation means (delta-beta value). More advanced methods seek to identify continuous differentially methylated regions (DMRs), which are biologically more informative. DMRs can be inferred both from microarray and bisulfite sequencing data. 

A number of tools have been developed by the scientific community for the purpose of identifying DMRs. Some of these include Rocker-Meth \cite{benelli2021charting}, which utilizes hidden Markov Models, DMRcate \cite{peters2021calling}, which employs kernel smoothing on single site t-statistics and Metilene \cite{juhling2016metilene}, which relies on a circular binary segmentation algorithm combined with a two-dimensional Kolmogorov–Smirnov test. These tools output a set of regions that are either hypermethylated or hypomethylated in one group w.r.t  the other. 

The analysis gets even more complex in a scenario with many groups, such as the one discussed in this work. For example, when analyzing methylomes of many different cell types, the goal is usually to find uniquely (de)methylated regions for each cell type. The most common method in literature is selecting top n regions/sites with the highest delta-beta in a one-cell-type--vs--all-other comparison \cite{loyfer2023dna, teschendorff2017comparison, van2021targeted}, or top n features of each class from a multi-class machine learning model \cite{chakravarthy2018pan, tang2018tumor}. While these markers might have discriminative power when combined, there is no guarantee they are truly unique. Therefore, in a setting where the goal is to examine the number of truly unique methylation markers of each cell type and their functions, the mentioned approach might not be optimal. I will address this problem in the following chapters by presenting a new, all-against-all method for identifying cell-type-specific markers.

\section{Resolving cell type mixtures using DNA methylation}

The problem of inferring in what amounts different cell types are present in a composite mixed sample is referred to as deconvolution. Although cell types can be purified from bulk samples \textit{in vitro} through methods like cell enrichment or cell sorting, this process is laborious and requires specific antibodies for each cell type of interest. A post-experimental approach to cell subpopulation analysis involves the use of computational deconvolution as an alternative method. To analyze cell-type mixtures accurately in such a way, the deconvolution software needs to be modeled on omics data with strong underlying cell-type-specific signatures, such as DNA methylation. Computational methylation-based deconvolution is hereinafter referred to as deconvolution.
Deconvolution models can operate both on CpG sites or larger regions. These methods can be divided into two main categories depending on whether they are using referenced, pure cell type data or not \cite{jeong2022systematic}:

\textbf{Reference-based} approaches infer cell type composition based on reference data of purified cell populations or single-cell experiments \cite{jeong2022systematic}. These methods consist of two main steps. Firstly, regions (or sites) with discriminative, cell-type-specific methylation need to be selected. Secondly, various algorithms and statistical models are utilized to fit cell type proportions using the observed methylation data in the selected features. Linear least squares methods \cite{loyfer2023dna, arneson2020methylresolver, schmidt2020deconvolution}, expectation–maximization algorithms \cite{zhang2021emeth, caggiano2021comprehensive}, deep neural networks \cite{levy2020methylnet}, various forms of regression \cite{chakravarthy2018pan, tang2018tumor, houseman2012dna} and others have been applied.

\textbf{Reference-free} methods, unlike reference-based approaches, utilize alternative strategies to estimate the proportions of various cell types without relying on reference data. Implementations are quite variable, and numerous such approaches have been proposed \cite{lutsik2017medecom,fong2021determining,lee2019prism,zheng2014methylpurify}. 

Both reference and reference-free methods are well established and have been used widely to infer cell populations in whole blood, cfDNA, and bulk tissues, including tumors.


These approaches have been extensively employed on microarray data, but nowadays on bisulfite sequencing data as well, which is becoming more popular due to reducing costs, single-nucleotide resolution methylation calls, and genome-wide coverage. Bisulfite sequencing also opens up the possibility of analyzing the methylation states of each DNA fragment (read) in the sample. Therefore, another division of deconvolution methods can be introduced:

\textbf{Beta-based} methods, which are applicable both on microarray and sequencing data, perform deconvolution by fitting cell type proportions using 0-to-1 ranged beta values. Microarrays already provide data in this format. However, sequencing reads have to be collapsed and summarized to express the fraction of methylation on each site or region, which might cause a loss of information. 

\textbf{Read-based} methods, which are only applicable to sequencing data, perform deconvolution by analyzing the methylation status of each read or read pair. This should theoretically be more suitable for deconvolution, given that the sample being deconvolved is a mixture of DNA fragments originating from different cell types. Recent scientific deconvolution projects seem to prefer this approach \cite{scott2020identification, loyfer2023dna, keukeleire2022cell}. However, an extensive comparison between read-based and beta-based methods, where all other variable factors have been eliminated, has still not been performed. 

In this research project, I will introduce a novel approach for feature selection in reference-based deconvolution and compare the performance of both beta-based and read-based deconvolution on an identical set of regions.

%% file: chapters/Results/results.tex
\section{Cell-type-specific methylation signals via pairwise comparisons}

\subsection{Pairwise DMR identification}

To identify the most suitable way of selecting discriminative regions to be used for composite tissue deconvolution, the uniqueness of cell type methylation markers needs to be assessed first. Instead of performing a one-vs-all analysis, I resort to pairwise comparisons later to identify the most informative markers and analyze their functional roles.  
From cell-sorted, WGBS data of healthy adults produced by \textit{Loyfer et al.} \cite{loyfer2023dna}, autosomal chromosomes were selected from 186 samples which I grouped into 44 cell types while ensuring all groups had at least three biological replicates. Considering only CpG sites, all groups had mean sequencing coverage above 20x, providing high granularity for the beta signal (Figure \ref{fig:coverage}).

\begin{figure}[ht!]
    \centering
    \hspace{-1em}\includegraphics[width=0.8\textwidth]{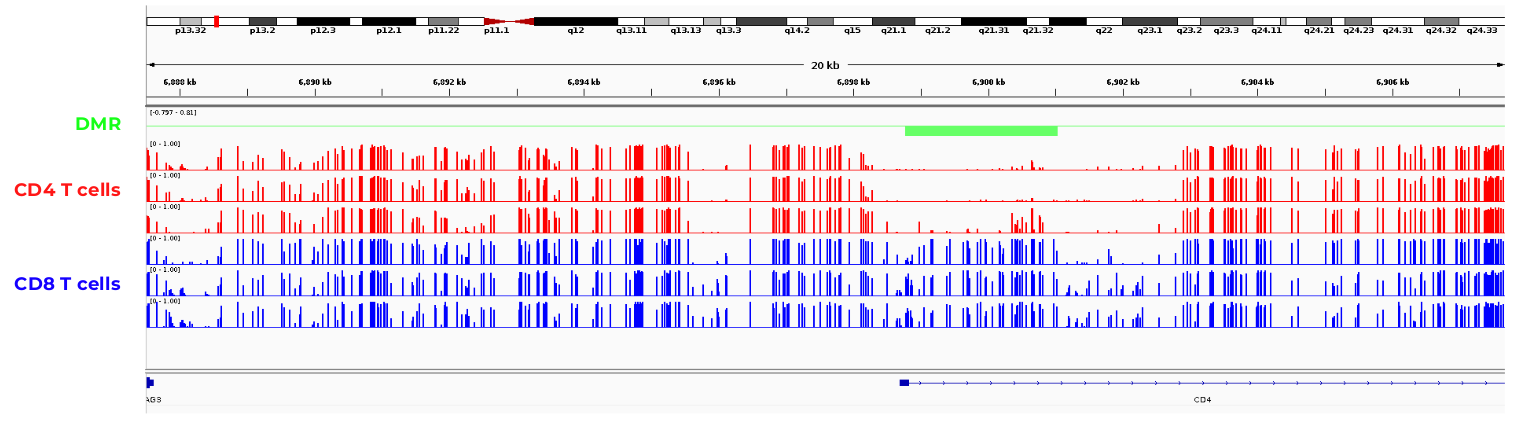}
    \caption{Differential methylation between CD4 and CD8 T cells samples captured by Metilene. Visualisation was made by using IGV \cite{thorvaldsdottir2013integrative}.}
    \label{fig:cd4cd8}
\end{figure}

Using Metilene \cite{juhling2016metilene}, a tool for identifying differentially methylated regions between groups, DMRs were called in all possible 946 pair-wise combinations between the 44 cell type groups. DMRs with at least 5 CpGs, FDR less or equal to 0.05, and a delta-beta value greater than 0.4 were retained in all comparisons. Figure \ref{fig:cd4cd8} shows an example of DMR defined between two groups of samples. Even in the case of closely related cell types, such as CD4 and CD8 T cells, Metilene can capture differential methylation. Here, the marked region is in line with the biological difference between these cell types -- the expression of the \textit{CD4} gene. Demethylation of the first exon and intron in \textit{CD4} gene suggests that DNA methylation is one of the mechanisms for regulating its expression.

\vspace{1em}

\begin{figure}[ht!]
    \hspace{-3em}\includegraphics[width=1.15\textwidth]{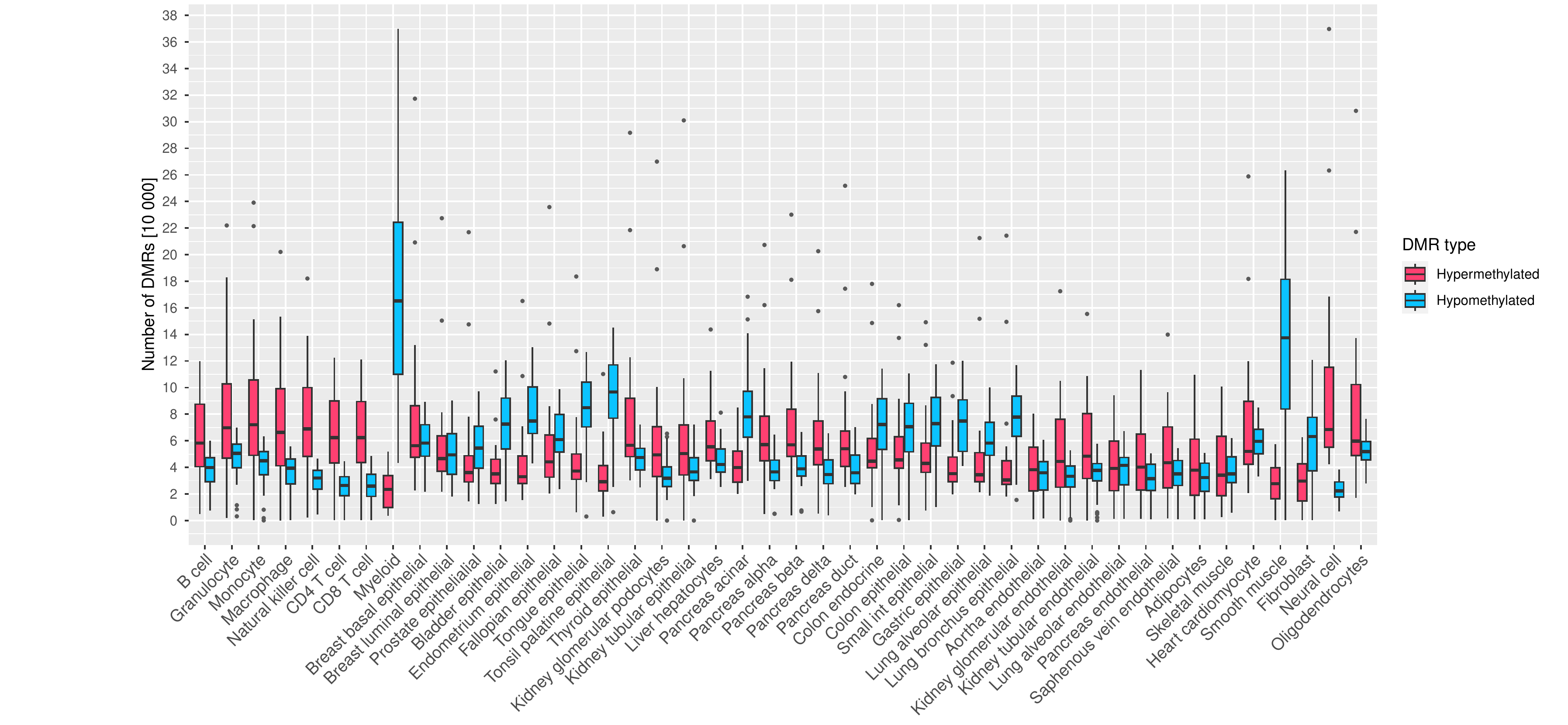}
    \caption{Number of hyper/hypomethylated regions w.r.t. all other cell types (expressed in 10.000s). }
    \label{fig:dmr_wrt_all_others}
\end{figure}

The number of DMRs identified w.r.t. all other cell types is relatively stable (Figure \ref{fig:dmr_wrt_all_others}), with the exception of myeloid and smooth muscle cells, which are heavily hypomethylated compared to all others. Myeloid cells are hematopoietic; thus, like other multipotent stem cells are expected to have more genome-wide demethylations \cite{cheng2015dna}. However, for smooth muscles, there is no clear explanation for this trend, nor has relative demethylation of this cell type been reported previously.
\vspace{1em}

\begin{figure}[ht!]
    \hspace{-2.75em}\includegraphics[width=1.1\textwidth]{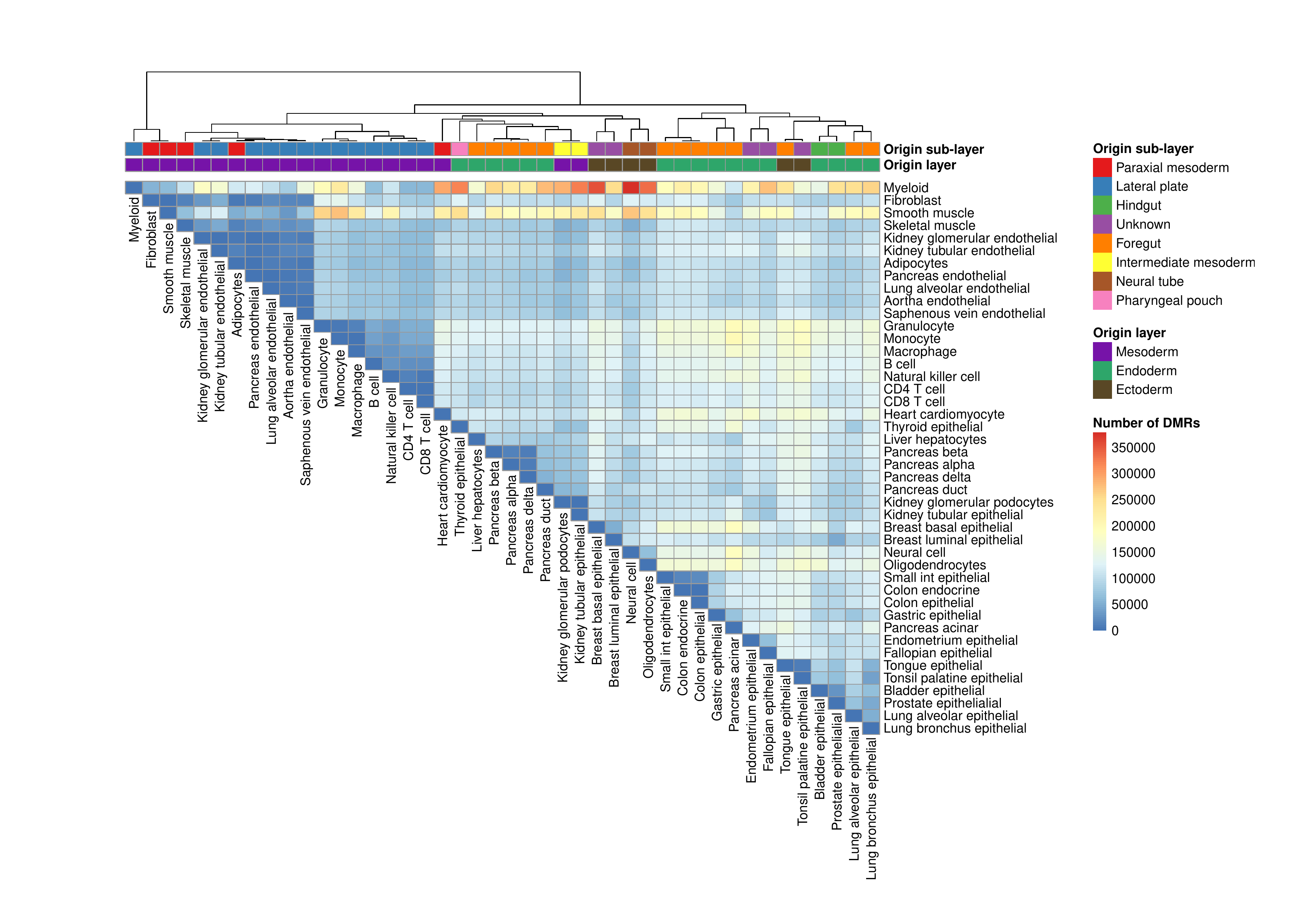}
    \vspace{-2.2em}
    \caption{Using the number of DMRs as a pseudo-distance for clustering with complete linkage.}
    \label{fig:dmr_distance_heatmap}
\end{figure}

The number of DMRs between two cell types can be used as a pseudo-distance to perform clustering (Figure \ref{fig:dmr_distance_heatmap}). In line with previous results, myeloid and smooth muscle cells have the highest degree of differential methylation across all cell types.
Interestingly, clustering into developmental layers and, in some cases even sub-layers, is visible. This indicates that methylation patterns are shared between cell types in the same developmental lineage and are preserved long after differentiating from multipotent stem cells. Likewise, the functional convergence of different cell types might explain this observation. Blocks of highly developmentally and functionally related cell types can be seen for endothelial, blood and various epithelial cell types. Most mesodermal cell types were grouped into a cluster very early, while other cell types formed smaller functional groups, such as gastrointestinal and pancreas tissues.
An interesting observation from these results is the proximity of adipocytes to endothelial cells. Methylation-wise, adipocytes and endothelia are even more similar to each other in comparison to the mutual resemblance of different types of white blood cells. 
Even though there is a strong preservation of developmental lineage signal across cell types, these results show that there are cases when tissues with completely different biological functions can converge to having a very similar methylome, thus suggesting the importance of understanding other epigenetic regulation mechanisms.

For the next analyses, cell types with too few DMRs between them were merged into a single category, and Metilene was re-run with aggregated groups. Kidney podocytes and kidney epithelial cells  (1 DMR), kidney glomerular endothelia and tubular endothelia (16 DMRs), smooth muscles and fibroblast (331 DMRs), monocytes and macrophages (370 DMRs), colon endocrine and epithelia (370 DMRs), and CD4 and CD8 T cells (711 DMRs) were joined into single groups before extracting unique methylation signatures. This aggregation procedure allows discovering more robust and informative markers while sacrificing only a small fraction of cell type resolution.

\subsection{Extraction of unique methylation signatures}

\begin{figure}[ht!]
    \hspace{1.5em}\includegraphics[width=1\textwidth]{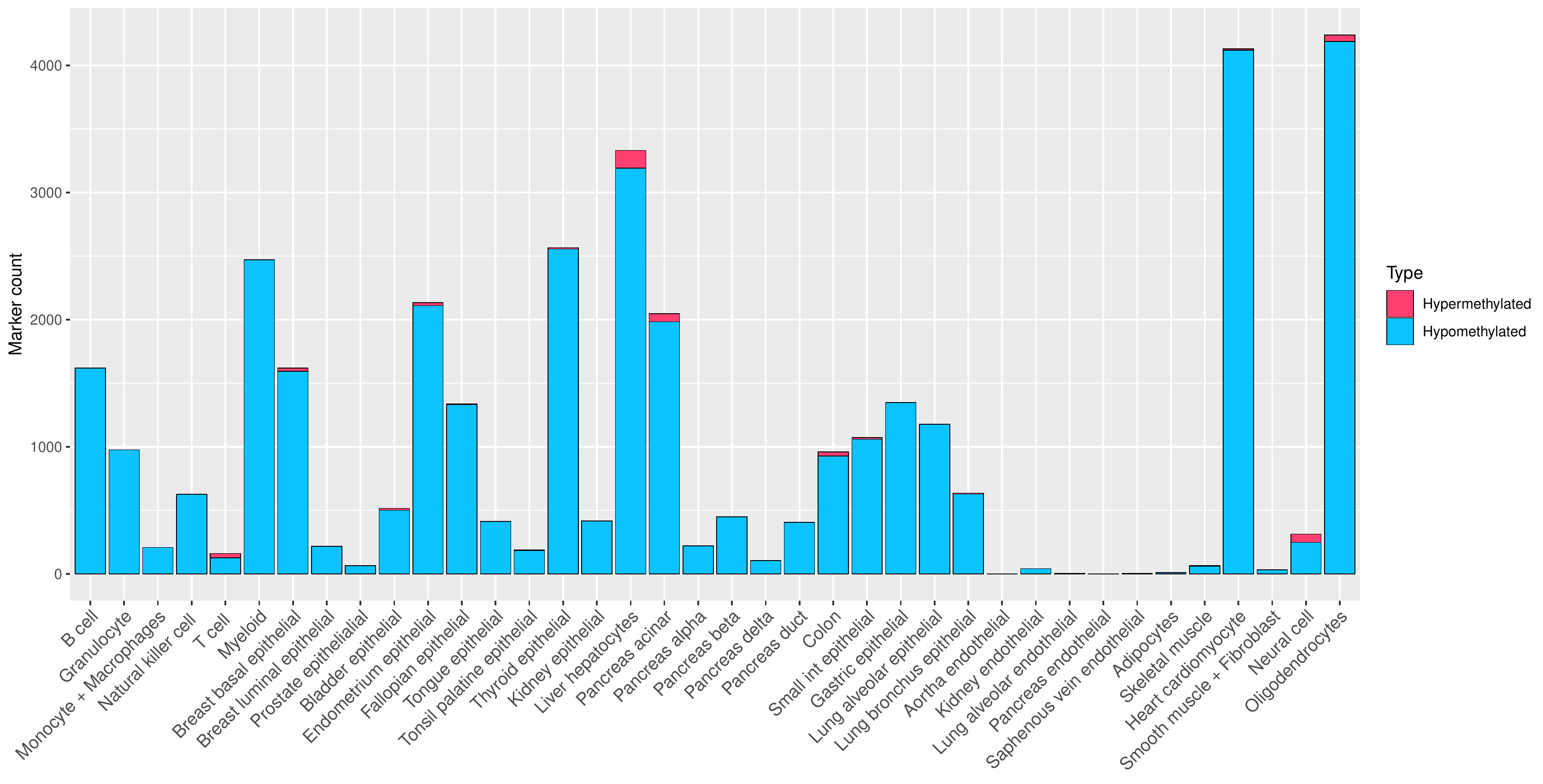}
    \caption{Number of hyper and hypomethylated UMRs across cell types.}
    \label{fig:marker_count_joined_ss_f}
\end{figure}

After obtaining all pairwise differential methylations, uniquely hypo and hypermethylated regions were identified for all 38 cell types. A region is defined as uniquely (de)methylated for a given cell type if it is identified as differentially methylated between that cell type and all others. This was performed separately for hypo and hypermethylated markers by intersecting all pairwise DMRs of a single cell type. Only regions with at least 3 CpG sites and an intra-group mean coverage of at least 10 were considered.
This approach identified a total of 35629 hypomethylated markers and 531 hypermethylated markers.
The number of uniquely methylated regions (UMRs) differs significantly across cell types, and hypermethylated unique regions are quite rare (Figure \ref{fig:marker_count_joined_ss_f}). Interestingly, smooth muscles (now grouped with fibroblast), which were heavily hypomethylated in pairwise comparisons (Figure \ref{fig:dmr_wrt_all_others}), have just a few UMRs. In order to exclude the possibility that aggregation with fibroblast confounded marker selection, pairwise DMR identification, and UMR selection were also performed when smooth muscle and fibroblast were separated and yielded a similarly low UMR count for both cell types (Figure \ref{fig:marker_count_sep}).
\vspace{1em}

Similarly, myeloid cells have fewer UMRs than hepatocytes, cardiomyocytes, and oligodendrocytes. This suggests that even though methylomes of these cells are globally demethylated w.r.t others, many of these hypomethylated regions are demethylated in at least one other cell type as well. In adipocytes and all endothelia, the number of UMRs is the lowest. 

\begin{figure}[ht!]
    \centering
    \begin{subfigure}[ht!]{\textwidth}
        \centering
        \includegraphics[width=0.75\textwidth]{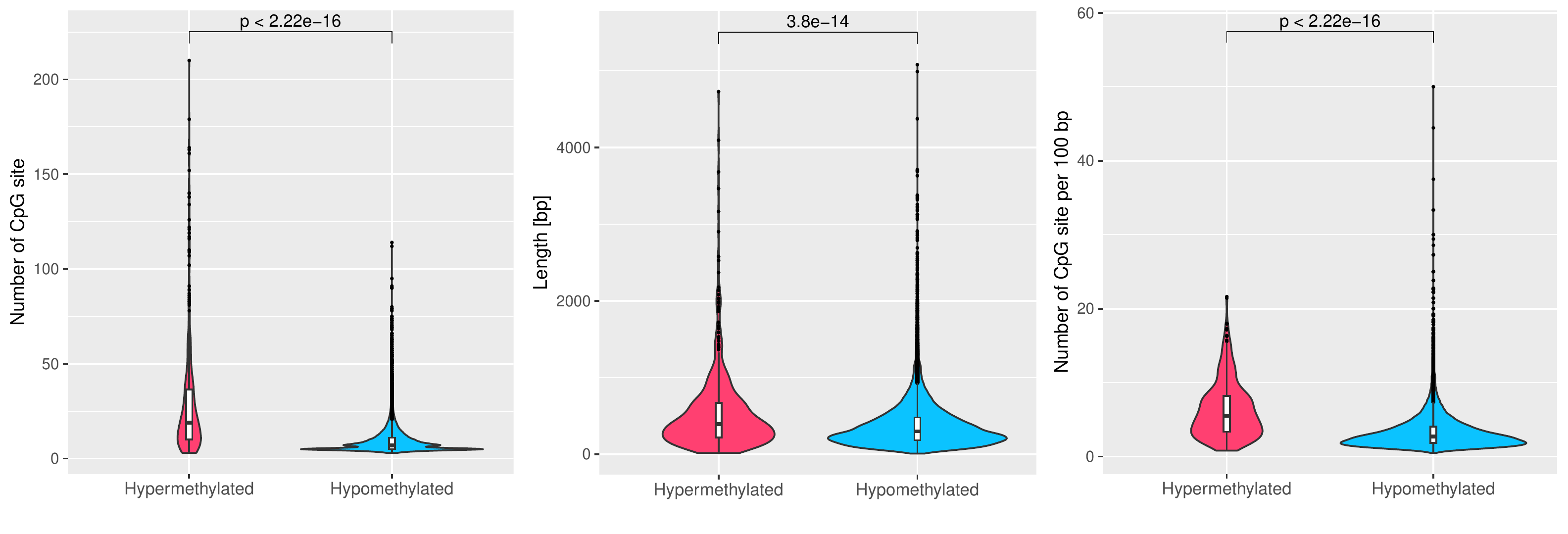}
        \label{subfig:global_props}
        \vspace{-0.5em}
        \caption{Properties of piled-up hypomethylated and hypermethylated UMRs.}
    \end{subfigure}
    
    \vspace*{1em}
    \begin{subfigure}{\textwidth}
        \hspace{-3.2em}
        \includegraphics[width=1.15\textwidth]{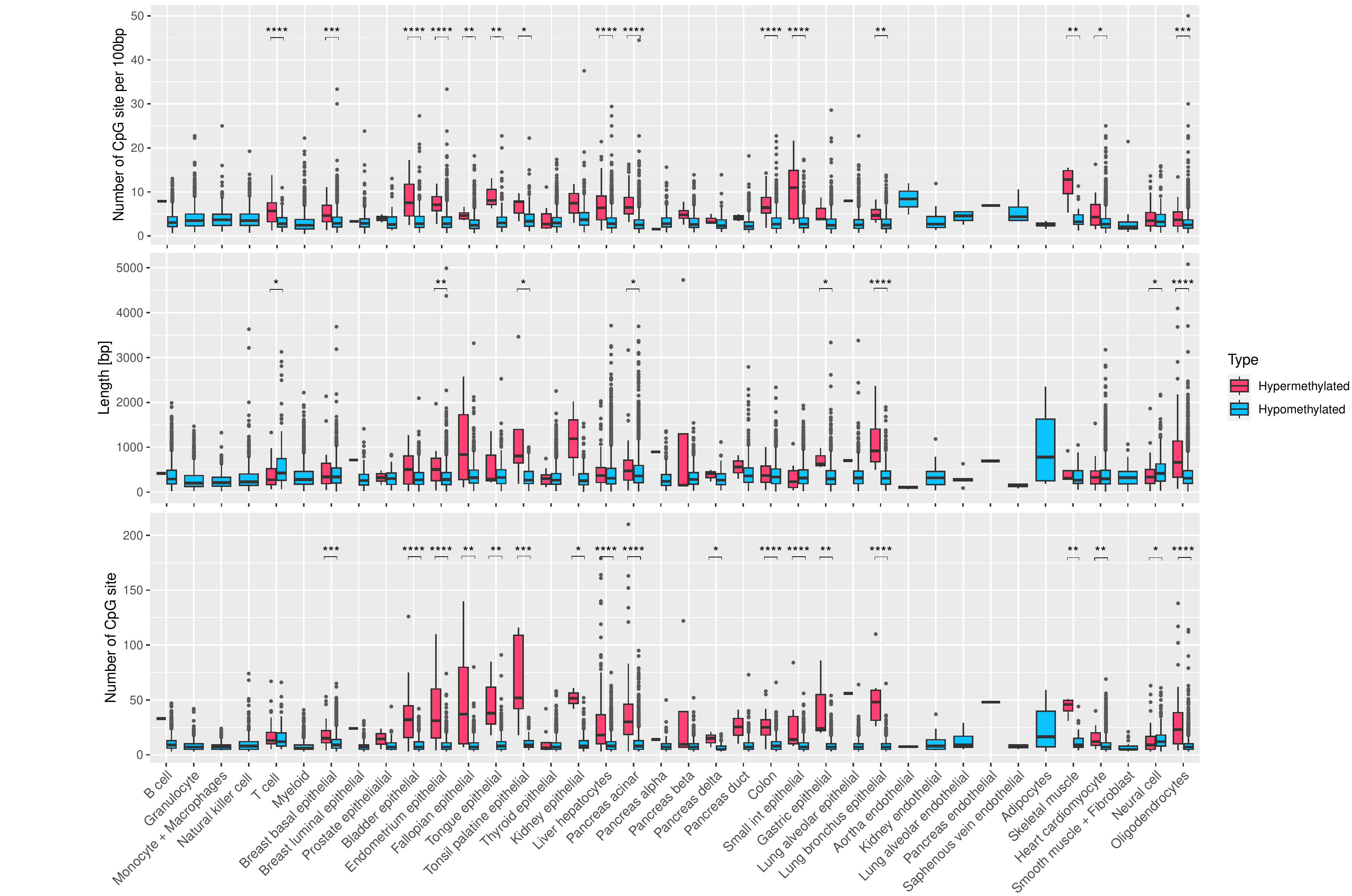}
        \label{subfig:cell_type_spec_props}
        \vspace{-1.25em}
        \caption{Properties of hypomethylated and hypermethylated UMRs across cell types.}
    \end{subfigure}

    \caption{CpG count, length, and density of hypomethylated and hypermethylated UMRs.}
    \label{fig:marker_info}
\end{figure}

Even though they are much less numerous when piled up, hypermethylated UMRs are significantly richer in CpG sites, longer, and have a higher CpG density (Figure \ref{fig:marker_info}) w.r.t. hypomethylated ones. This is also true at the cell type level in many cases.
As a sanity check, beta values of UMRs are shown in Figure \ref{fig:beta} to make sure hypomethylated markers have low and hypermethylated markers have high values. Comparing the set of all UMRs with the top 1000 markers of 36 different cell types, identified by \textit{Loyfer et al.}, showed that only 10570 are shared (difference in cell types due to different merging choices). To better understand the biological relevance of these markers, UMRs were investigated by assessing their functional properties using multiple approaches and by integrating them with different omics data.

\hspace{7em}

\subsection{Distribution of functional elements covered by UMRs}
An important factor to consider is which functional elements can the identified markers be found in. Since DNA methylation is most commonly associated with CpG islands, the presence of markers inside these regions was assessed first. Surprisingly, out of 36160 UMRs identified only 728 overlap CpG islands, 292 of which are hypermethylated regions. This is quite a strong enrichment, given that hypermethylated UMRs constitute only $\sim$1.5\% of the total marker count. These results also suggest an explanation for the higher CpG numbers, density, and longer regions in hypermethylated w.r.t hypomethylated markers.

\begin{figure}[ht!]
    \centering
    \begin{subfigure}[ht!]{\textwidth}
        \centering
        \hspace{2em}\includegraphics[width=0.9\textwidth]{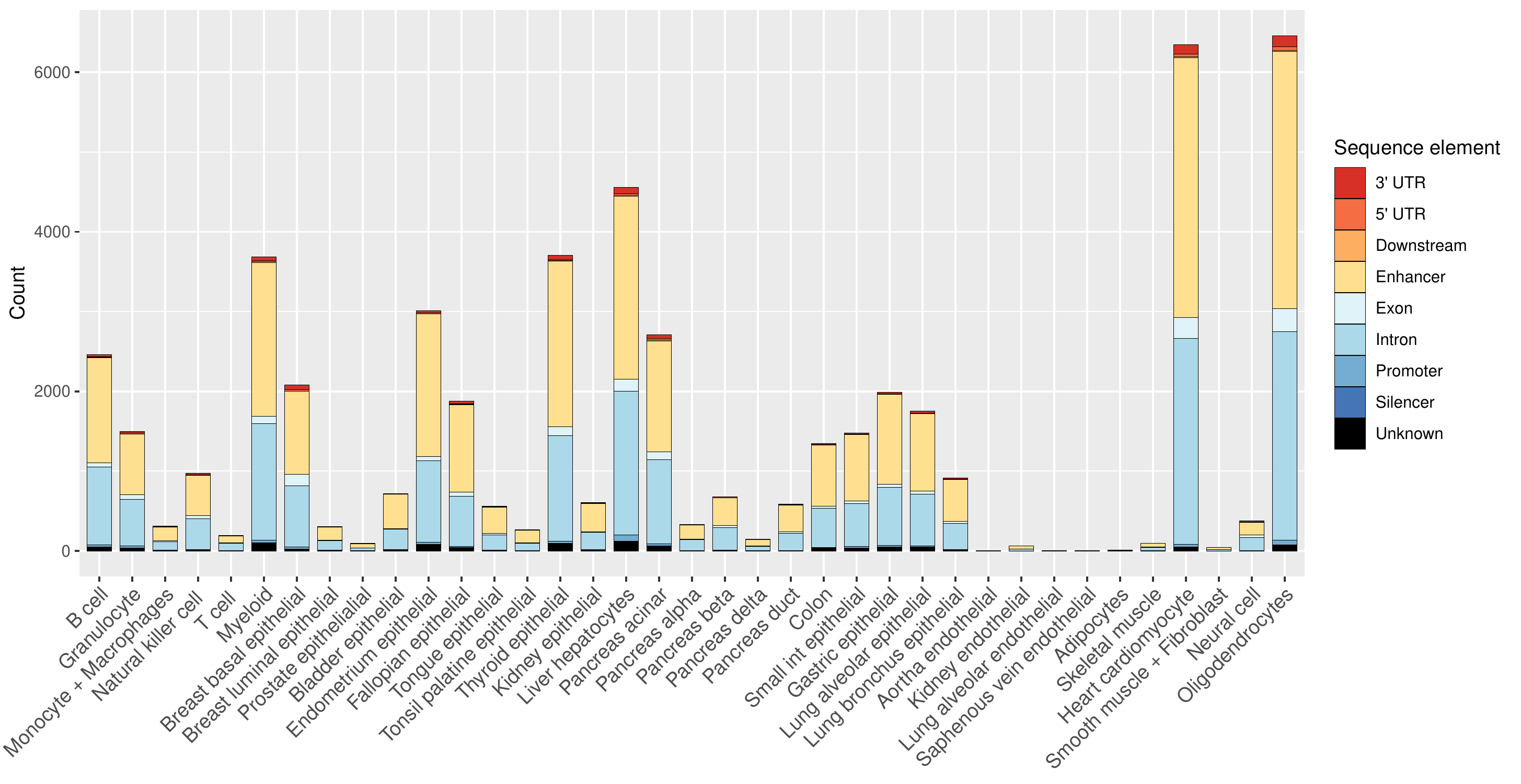}
        \label{subfig:hypo_functional}
        \caption{Hypomethylated UMRs}
    \end{subfigure}
\vspace{2em}
    \begin{subfigure}[ht!]{\textwidth}
        \centering
        \hspace{2em}\includegraphics[width=0.9\textwidth]{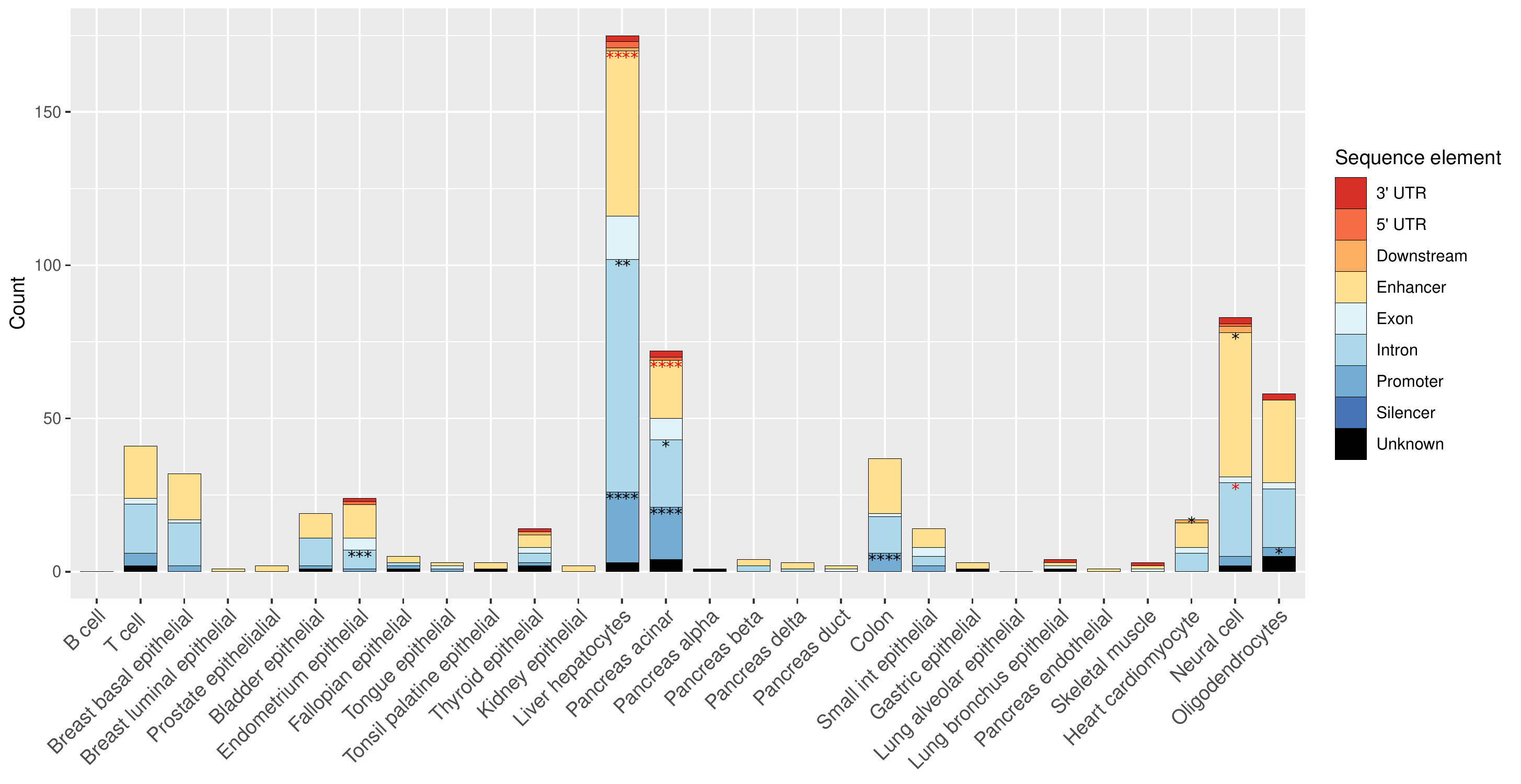}
        \label{subfig:hyper_functionl}
        \caption{Hypermethylated UMRs -- black asterisks denote the significance of enrichment, while the red ones depletion w.r.t hypomethylated markers of the same cell type.}
    \end{subfigure}
    \vspace{-2em}
    \caption{Functional elements overlapping cell-type-specific markers.}
    \label{fig:hyper_hypo_functional}
\end{figure}

Next, I investigated which functional elements overlap UMRs by matching markers with various gene annotations (Figure \ref{fig:hyper_hypo_functional}). Apart from using standard UCSC gene annotation data, I also included experimentally validated, multi-cell-type databases of 1988217 enhancers collected by \textit{Gao et al.} \cite{gao2020enhanceratlas} and 8827 silencers collected by \textit{Zeng et al.} \cite{zeng2021silencerdb}. The distribution of functional elements across cell types, both in hypo and hypermethylated UMRs, is quite consistent -- markers are mostly found in enhancers and introns. These results are consistent with previous studies that have associated cell-type-specific demethylation with gene enhancers \cite{loyfer2023dna, dor2018principles, kirillov1996role}. When aggregated and compared, hypermethylated UMRs are significantly enriched for promoters and exons (Fisher's exact test p-value = 1e-44 and 4.56e-6, respectively). In contrast, hypomethylated ones are enriched for enhancers (Fisher's exact test p-value = 5.37e-10). 

\vspace{2em}

\subsection{Exploring potential functions and cell-type-specificity of UMRs using gene and motif enrichment}

To evaluate whether selected UMRs support cell-type-specific characteristics and assess their joint functional effect across cell types, genomic region enrichment was performed using rGREAT \cite{gu2023rgreat} to identify genes adjacent to markers and analyze their enrichment for various GO annotations. Selected significant results for hypomethylated gene enrichment are shown in Table \ref{table:gene_enrichment}.  

\vspace{1em}
\begin{table}[ht!]
    \hspace{-2.5em}
    \includegraphics[width=1.1\textwidth]{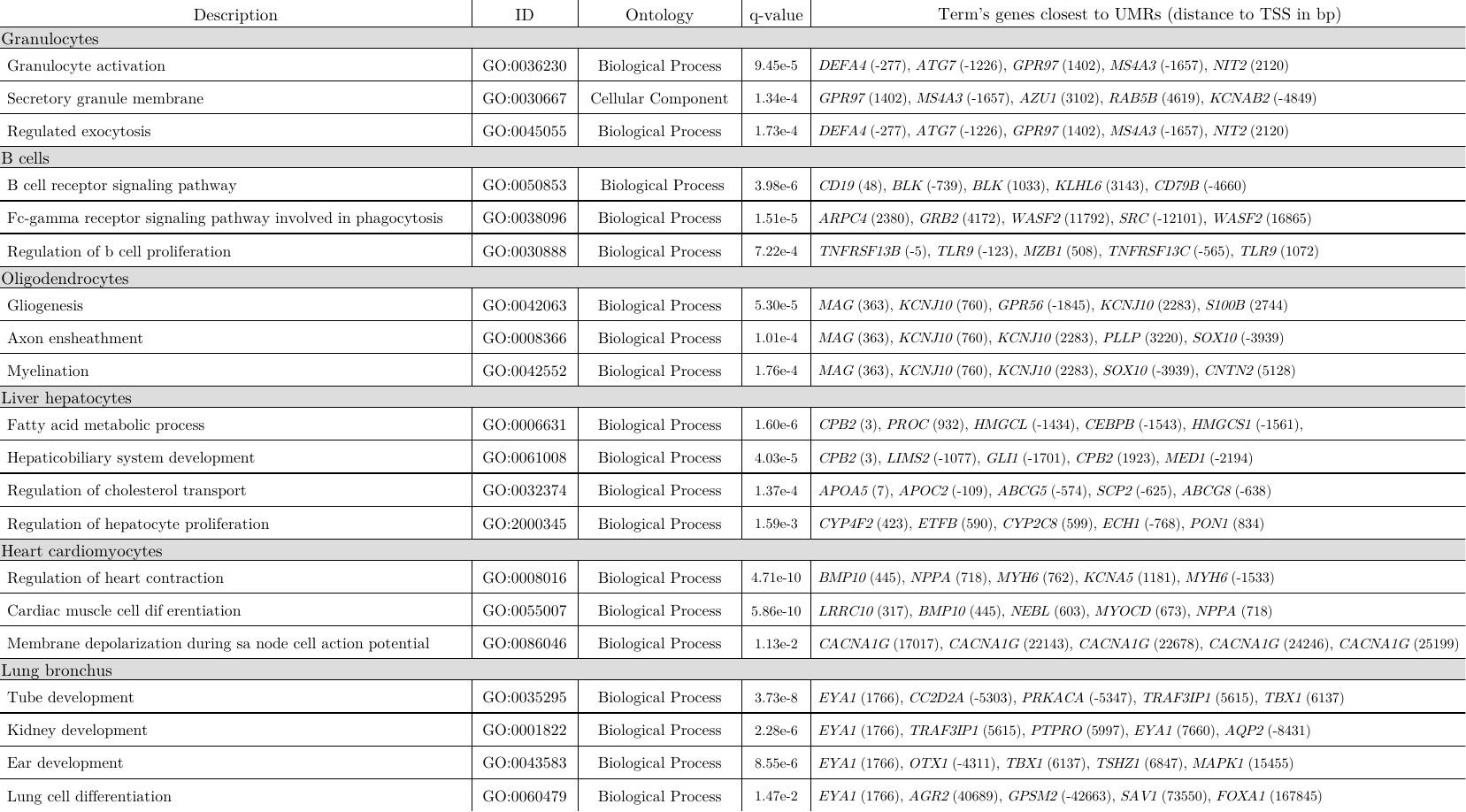}
    \caption{Gene enrichment results with top five genes of a given term based on the distance of transcription start site (TSS) to the closest hypomethylated UMR. Negative distance denotes that the position of the element is upstream of TSS.}
    \label{table:gene_enrichment}
\end{table}

Blood cells show enrichment for terms related to their physiologies; oligodendrocytes terms related to myelination, which is their primary function; liver hepatocytes are enriched in terms related to fat processing and cardiomyocytes in functions that characterize heart muscles uniquely. Interestingly, besides being enriched for lung cell terms, bronchus cells are enriched for several terms related to tubulogenesis. Similar gene pathways are known to be responsible for tubular growth and branching both in bronchial and nephronic zones \cite{quaggin1999basic}. The same could be possible for ear development.

Hypermethylated UMRs around gene loci, which are generally correlated with reduced activity, also show interesting results. In all cell types with identified hypermethylated UMR enrichment results, terms related to fetal development are present. Lung bronchus epithelia were enriched for pattern specification processes and respiratory system development (q-value = 0.023; 0.046, respectively). Hepatocytes were enriched for developmental growth (q -value = 0.016). Oligodendrocytes were enriched for nervous system development (q-value = 4.02-e4). In line with previous research, these results suggest DNA that methylation might be one of the mechanisms responsible for inhibiting pathways that should be active only during embryonic/fetal development \cite{koukoura2012dna}. 
~\\
Next, I aggregated all UMRs, separating hyper and hypomethylated ones, and performed motif enrichment using HOMER \cite{heinz2010simple}. In hypermethylated UMRs, the only significantly enriched motif was the binding site of the CTCF transcriptional repressor, identified in 5.46\% sequences. This protein is known to cause chromatin looping and genome folding by binding to unmethylated regions and acting as an insulator \cite{sun2022ctcf, wang2012widespread}. Conversely, hypermethylated UMRs overlapping CpG islands showed no significant motif enrichment.

A wide range of motifs in low percentages of target sequences was identified in hypomethylated UMRs, but no significant hits were identified in regions overlapping CpG islands. To better understand these results, motif enrichment of hypomethylated UMRs was run for each cell type individually. Table \ref{table:motifs} shows the selected enriched motifs of several cell types. In line with gene enrichment, a high cell-type-specificity is observed in the results, and transcription factors related to correct cell types have been identified. Interestingly, bronchus cells are enriched for a nephron progenitor transcription factor binding site, providing more evidence of the developmental similarity between bronchus and kidney tissues.
In basal breast epithelia, 47\% of UMRs had a HIC1 binding site. This transcriptional repressor is a tumor suppressor often found hypermethylated in cancer (its name is an abbreviation for Hypermethylad In Cancer 1). Upon investigating which genes are located near UMRs with HIC1 binding sites, gene groups previously reported to interact with HIC1 were identified \cite{zheng2012signification}. These include UMRs observed in the first intron of the fibroblast growth factor 2 (\textit{FGF2}) gene, the second intron of fibroblast growth factor receptor-like 1 (\textit{FGFRL1}), and 10kb upstream to C-X-C motif chemokine receptor 1 (\textit{CXCR1}) gene. However, many genes with previously unreported HIC1 interaction have also been identified. UMRs with HIC1 binding sites were observed in TSSs of interleukin 17B (\textit{IL17B}) and persephin (\textit{PSPN}) genes, and in the first exons of keratin 6B (\textit{KRT6B}) and \textit{DIABLO} (an inhibitor of apoptosis) genes.
In prostate cells, HOXB13 transcription factor linked to prostate development and recurrence following radical prostatectomy was observed in $\sim$60\% of hypomethylated UMRs. Upon identifying genes in the proximity of UMRs with this binding site, only a single gene with previously reported HOXB13 interaction was identified -- a marker of proliferation Ki-67 gene (\textit{MKI67}) with a UMR in the 12\textsuperscript{th} intron \cite{yao2019homeobox}. Among other previously unreported genes, hypomethylated UMRs were identified in the first introns of \textit{DROER} (ERH mRNA splicing and mitosis factor) and \textit{ADH4} (alcohol dehydrogenase 7) genes. It is unknown whether DNA methylation affects the binding affinity of HIC1, however, studies have shown that HOXB13 prefers methylated binding sites \cite{yin2017impact}. As previous research has shown that overexpression of HOXB13 is related to prostate cancer recurrence \cite{zabalza2015hoxb13}, it might be possible that this is accompanied by hypermethylation of its targets. This also underlines the versatility of DNA methylation in the cell-type-specific context. It is not only associated with reducing the binding affinity of transcription factors, but also its absence can cause the same effect. Anyhow, these results can provide a list of putative targets and identify new regulatory mechanisms in different cell types. 

\begin{table}[ht!]
    \centering
    \includegraphics[width=0.9\textwidth]{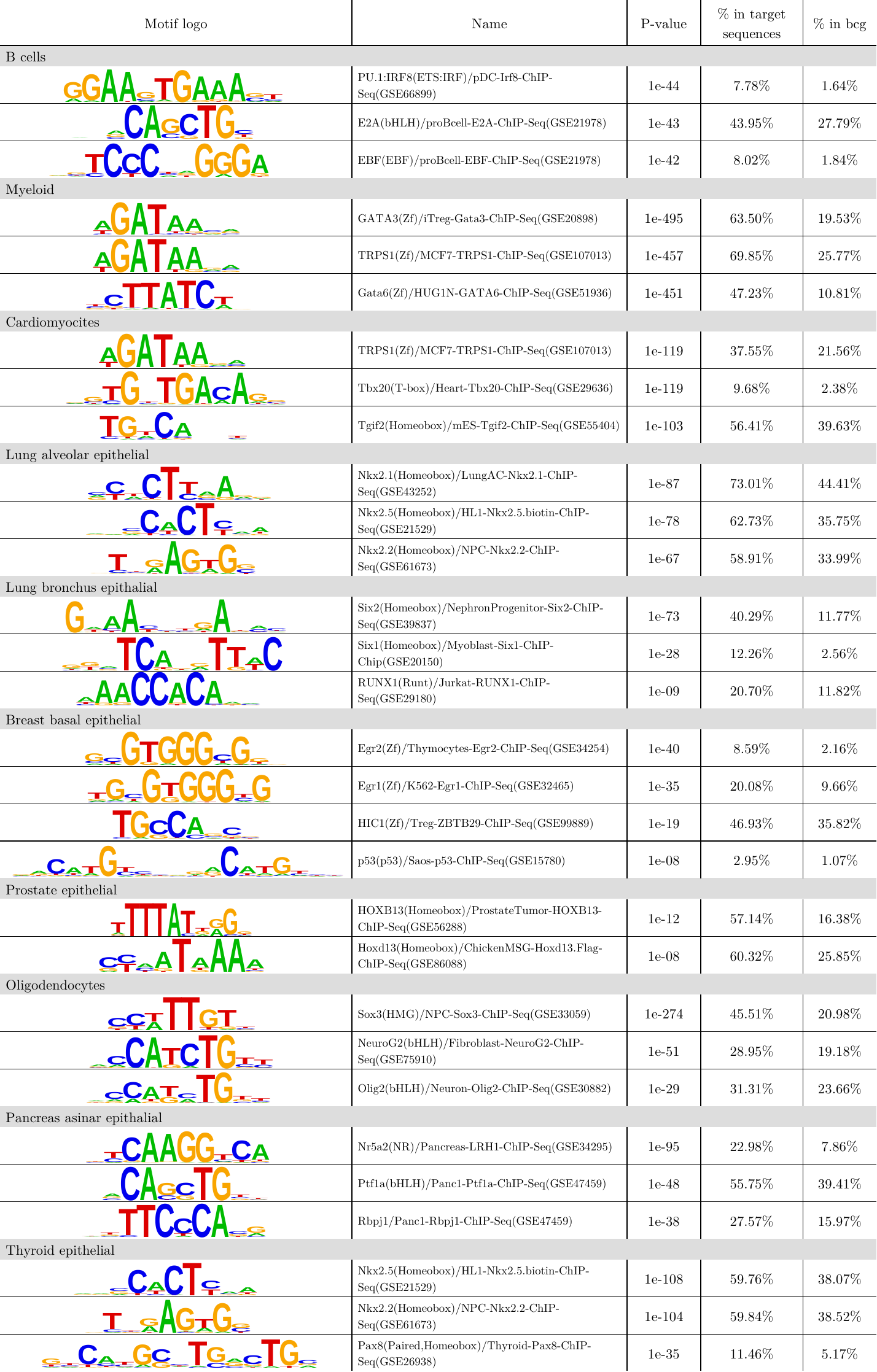}
    \caption{Motif enrichment results using HOMER.}
    \label{table:motifs}
\end{table}

For all motif and gene enrichment results and complete lists of gene-annotated UMRs with HIC1 and HOXB13 binding sites in basal breast and prostate epithelia, refer to the \textit{Data availability} chapter.

\subsection{Chromatin states in hypomethylated marker regions}

To assess the relationship between cell-type-specific methylation and chromatin modifications, hypomethylated UMRs were compared with 15-state ChromHMM Roadmap genome annotations \cite{roadmap2015integrative} for several cell types. 15-state ChromHMM is a bioinformatic resource that integrates epigenetic data of five histone modifications and annotates genomic regions with one of 15 chromatin states based on different combinations of histone marks present at that site. These chromatin states correspond to predicted but not verified functional elements. 

\begin{figure}[ht!]
    \hspace{-3em}
    \includegraphics[width=1.15\textwidth]{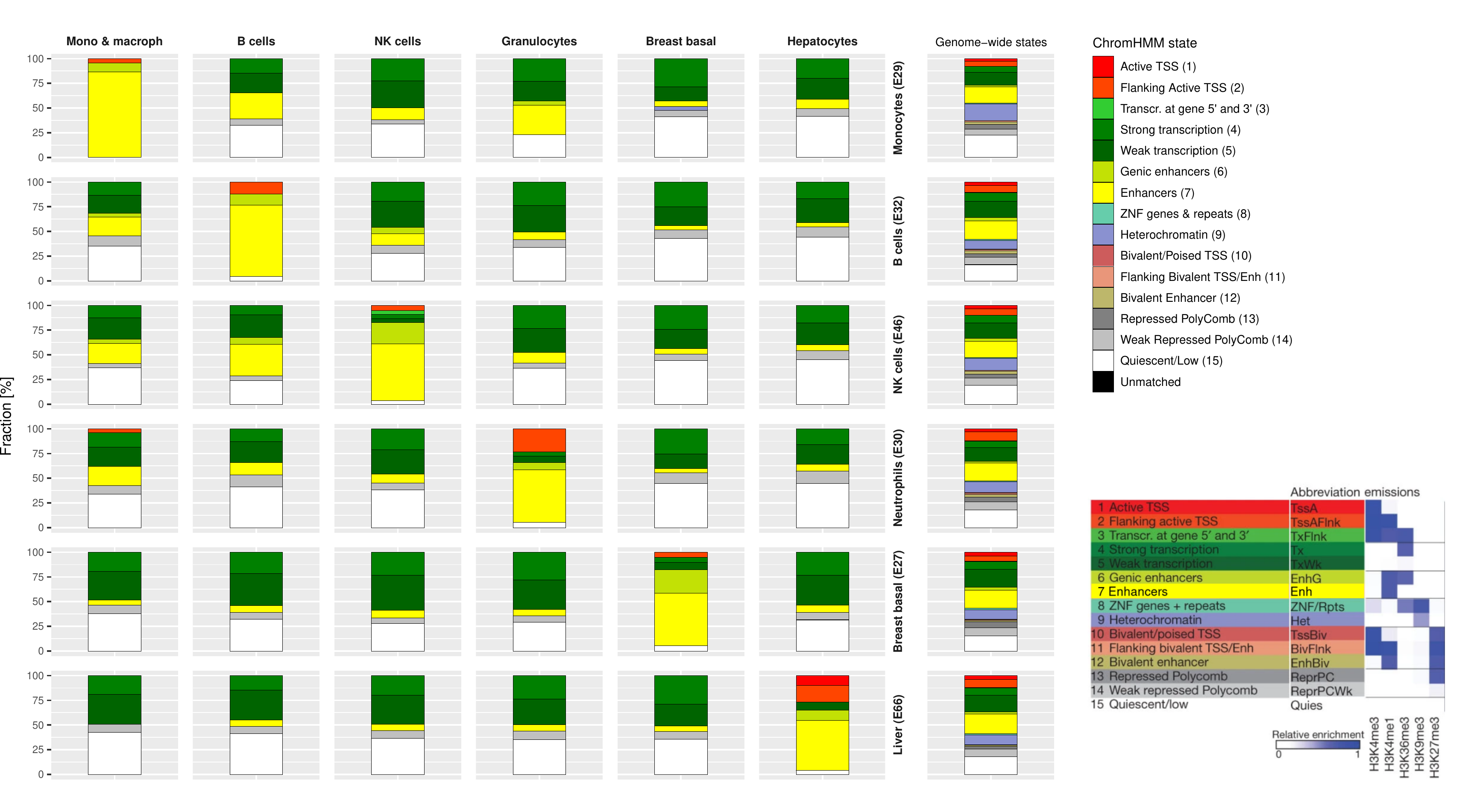}
    \caption{Fractions of chromatin states when overlapping hypomethylated UMRs with corresponding and non-corresponding ChromHMM annotations. The legend with the strength of chromatin marks was taken from \textit{Kundaje et al.} \cite{roadmap2015integrative}}
    \label{fig:chromHMM}
\end{figure}

As most of these state annotations are available for bulk tissues in the Roadmap project, I chose a few pure-cell samples and assessed their overlap with hypomethylated markers. All used annotations came from sorted cell assays apart from liver tissue, which was determined to be primarily composed of hepatocytes \cite{loyfer2023dna} and can be considered relatively pure. When a set of UMRs is matched with the corresponding cell type annotation (diagonal), a strong enrichment in regions predicted as enhancers (strong H3K4me1), genic enhancers (both strong H3K4me1 and H3K36me3), and active transcription start sites (both strong H3K4me3 and weak H3K4me1) can be observed (Figure \ref{fig:chromHMM}). In line with the results shown in Figure \ref{fig:hyper_hypo_functional}, most UMRs in matched annotations correspond to enhancer-like regions. However, when UMRs are matched with non-corresponding ChromHMM annotations (annotations for the same regions in another cell type -- outside the main diagonal), there is a significant enrichment in regions annotated as quiescent (no marks), weakly transcribed (weak H3K36me3), weak polycomb repressed (weak H3K27me3) and surprisingly, strongly transcribed (strong H3K36me3), possibly marking intronic enhancers. Non-diagonal proportions of these elements were compared to the diagonal ones using Fisher's exact test, and all yielded a p-value $<$ 2.2e-16.

\section{Deconvolving composite tissue samples via set cover region selection}

\begin{figure}[ht!]
    \hspace{-16em}\includegraphics[width=1.8\textwidth]{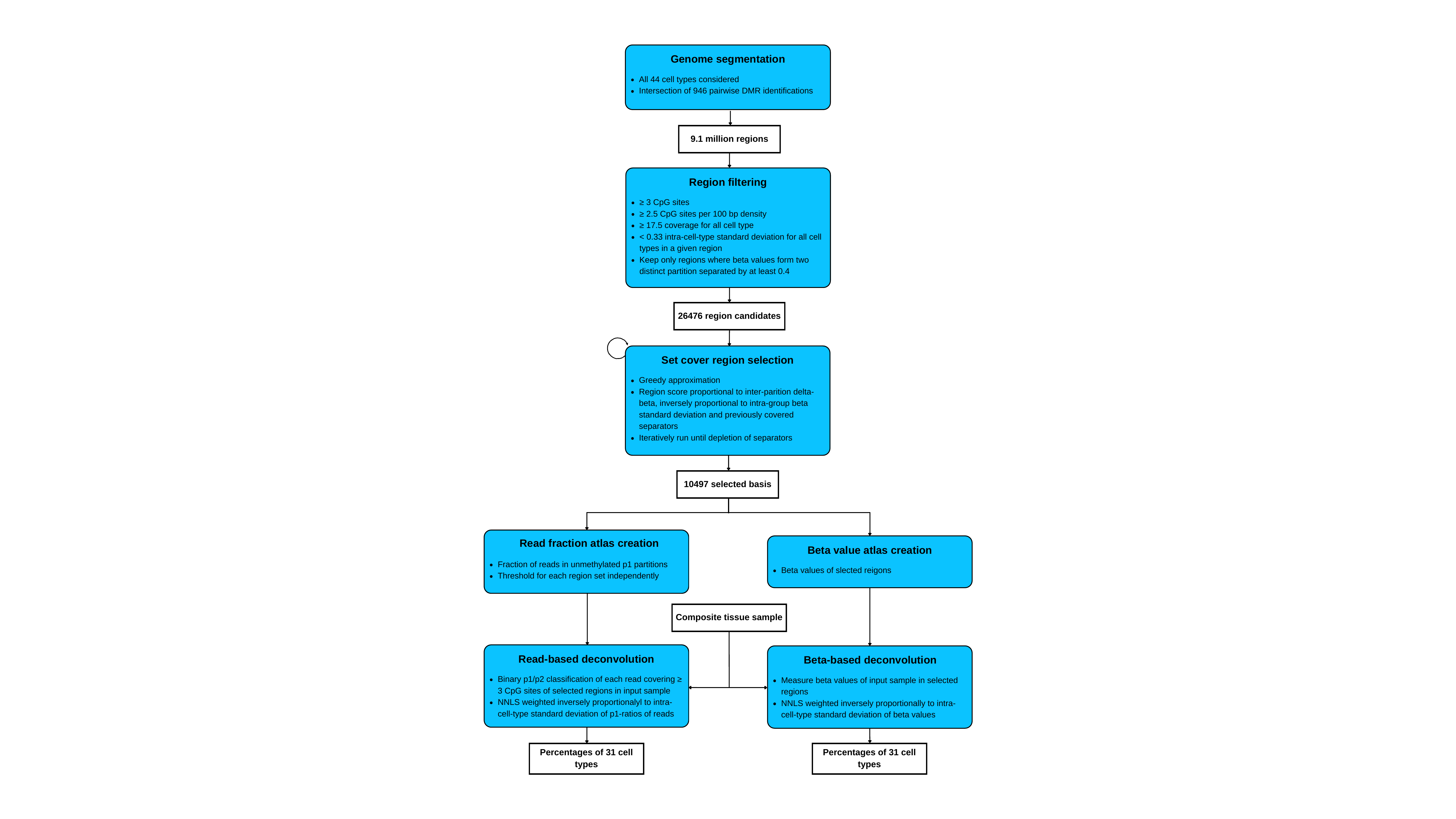}
    \vspace{-3em}
    \caption{Diagram showing each key step in the proposed deconvolution approach.}
    \label{fig:diagram}
\end{figure}

Apart from investigating cell-type-specific methylation, a secondary goal of the previous chapters was to determine how regions should be selected in a multi-cell-type deconvolution problem. Since the number of UMRs is very variable between different cell types and can be as low as one, these cannot be used for deconvolution, thus, another method for identifying discriminative regions should be applied. This chapter presents a deconvolution pipeline that utilizes a novel method to select regions based on pairwise methylome differences using the set cover algorithm combined with a weighted non-negative least squares optimization (NNLS). As no publication, to my knowledge, has rigorously compared the performance of read-based and beta-based deconvolution methods on the same set of regions and equal terms, this is an additional goal of this project. The overview of the proposed approach is shown in Figure \ref{fig:diagram}.

Firstly, a genome segmentation of informative regions was obtained by intersecting all 946 DMR identification files (Metilene output) for all 44 cell types -- no aggregation was performed \textit{a priori}. This resulted in around 9 million regions. 

Filtering  was performed based on the number of CpG sites -- at least 3; CpG density -- at least 2.5 CpG sites per 100 bp; coverage -- at least 17.5x for all cell types in a given region and intra-cell-type standard deviation -- standard deviation between samples of the same cell type has to be less than 0.33 for all cell types. These filtering thresholds were chosen because they provide a good trade-off between the number of separable cell types in the final selected set and keeping only highly-informative and trustable regions. Additionally, only regions where beta values of cell types form two distinct partitions (unmethylated - $p1$; methylated - $p2$), separated by at least 0.4, were selected. This kind of filtering enables the creation of binary vectors of pairs of cell types a region can separate -- only pairs where cell types are located in different beta value partitions are separable. As some cell types had very few high-quality regions able to separate them, they were joined into bigger groups. 1) CD4 and CD8 T cells; 2) monocytes and macrophages; 3) colon endocrine and epithelia; 4) kidney tubular and kidney podocyte endothelia and  5) all endothelial cell types, adipocytes, smooth and skeletal muscles were joined into single groups. A few regions where members of these groups fall into separate partitions were also removed not to introduce intra-group variability (See \textit{Methods and Resources}). These groups are hereinafter also referred to as cell types. The filtering procedure resulted in 26476 candidate regions from which the final set was selected.

\begin{figure}[ht!]
    \centering
    \hspace{4em}\includegraphics[width=0.8\textwidth]{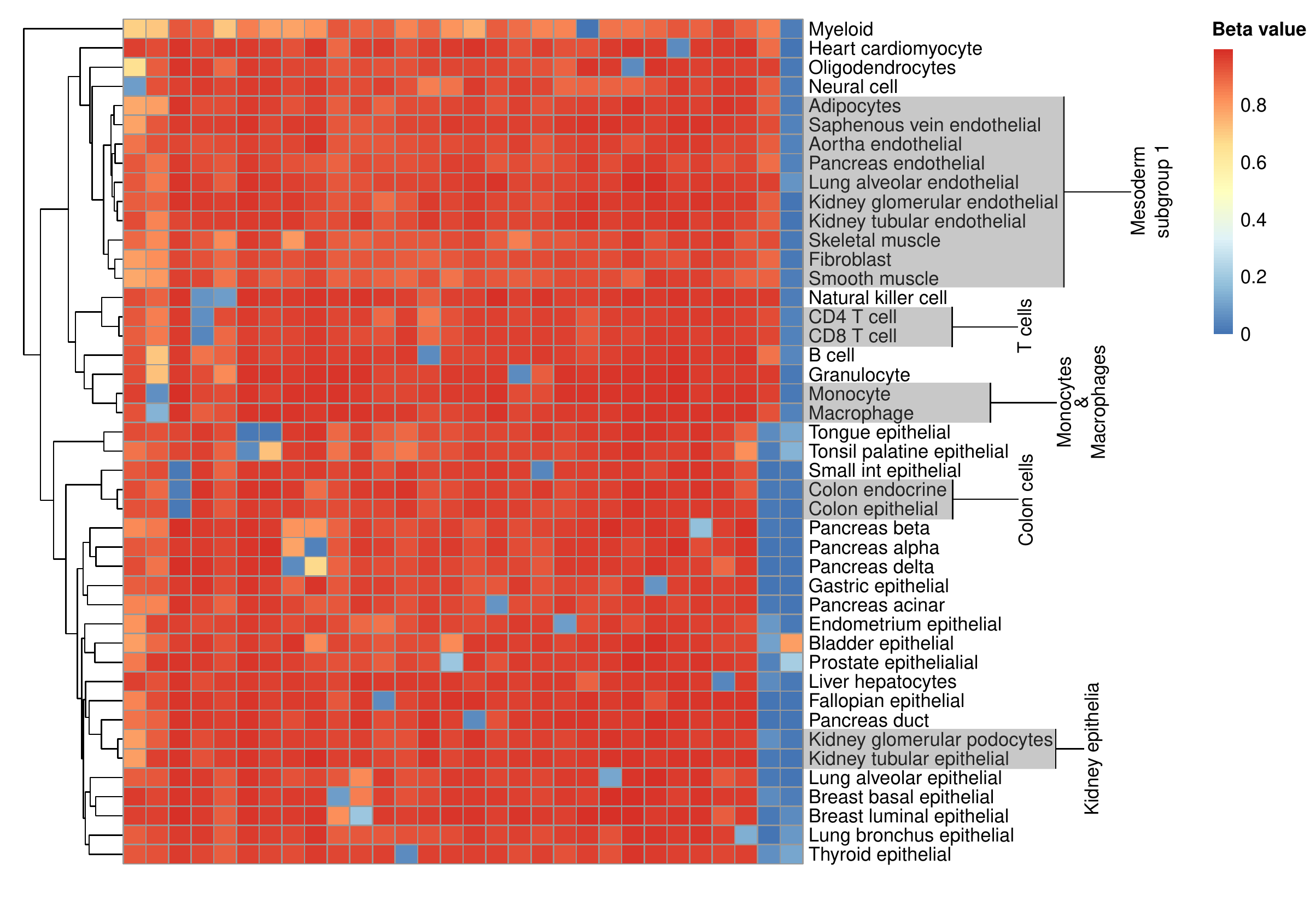}
    \caption{Beta values of 30 selected regions across all cell types. Clustering was performed using Manhattan distance and complete linkage. Joined cell types are shaded in gray.}
    \label{fig:single_basis}
\end{figure}

Next, a greedy approximation of the set cover algorithm was employed to identify minimal region subsets, using which all cell types are separable in terms of methylation. If we think of every region as a subset of cell-type separators it can provide, then collecting all separators means selecting the appropriate regions. Set cover is an NP-hard algorithm that provides a minimal region subset, where the union of all region separators covers all cell-type separators. Due to exponential computational complexity, it is not feasible to run this algorithm on many regions; therefore, a greedy approximation is used. 
In each round of the greedy set cover, the score of each region is weighted proportional to the inter-partition delta-beta (difference between inner boundaries of $p2$ and $p1$), inversely proportional to maximal intra-cell-type standard deviation, and inversely proportional to the sum of separators present both in the subset being built and the assessed region. This weighting scheme forces the algorithm to prefer selecting regions with high delta-beta, low intra-cell-type standard deviation, and those that maintain uniformity in the number of selected separators. Set cover was run iteratively 369 times, until depletion of any separator, while removing the selected regions from the candidate pool. After 369 runs, some separators became depleted; therefore, generating a region subset able to separate all cell types was impossible. Each of these 369 region subsets contained around 30 regions, and together they formed a set of 10479 regions that were used for deconvolution. Figure \ref{fig:single_basis} shows a subset of 30 regions from a single set cover run (Manhattan distance with complete linkage). Even with 30 regions, there are traces of biological clustering. Cell types joined into single groups do not have large variations of region beta values between them, since all regions separating them with a delta-beta of at least 0.4 were removed. Figure \ref{fig:all_basis} shows clustering on beta values of all 10479 selected regions. 

An atlas containing reference values for each cell type in each selected region has to be built before deconvolution can be performed. Two atlases, based on the same selected regions, were created separately for beta-based and read-based approaches. The beta-based atlas contains beta values of each region across all cell types. Atlas references of groups consisting of multiple inseparable cell types were created by calculating the mean beta values of all group members. For the read-based method, an atlas was created by determining the fractions of reads in the unmethylated partitions ($p1$). Moreover, each region has its own threshold for assigning reads to this partition: this is calculated as the midpoint between $p2$ partition's minimal and $p1$ partition's maximal value. Considering only read CpGs that fall inside a given region, if their mean methylation value is less than a threshold, it is assigned to the unmethylated partition $p1$; otherwise, to $p2$. Read-based atlas references of groups consisting of multiple inseparable cell types were created by calculating mean $p1$ read fractions.

To determine cell-type abundances in an input sample, methylation in selected regions is first measured based on reads or beta values. For each region, a linear equation is created where the unknown is the fraction of each cell type. Considering all regions this way, the result is an inconsistent, over-complete system of equations where many contradict each other. This problem is solved numerically using weighted NNLS. In the case of beta-based deconvolution, the weight is inversely proportional to the intra-cell-type standard deviation of beta values. In the case of read-based, it is inversely proportional to the intra-cell-type standard deviation of fractions of reads in chosen partitions.

\subsection{Simulations on synthetic cell mixtures}

To compare the performance of my proposed methods and assess their accuracy when deconvolving mixtures at different mixing ratios and coverage values, I used Blueprint \cite{fernandez2016blueprint} blood cell WGBS data to create synthetic mixtures. Selected Blueprint samples come from monocytes, T cells, B cells, and natural killer cells and can be considered relatively pure, as they are generated in sorted cell array experiments. Synthetic mixtures contain two cell types, one is taken as the minor fraction and the other as the background. Cell types were mixed at 5/95\%, 1/99\%, and 0.5/99.5\% ratios, at mean sequencing coverage values of 1.5, 5, 15, and 28x. Mixtures were generated for all possible combinations of input samples, mixing ratios, and sequencing coverage values -- this resulted in 144 simulated mixtures in total. 

Besides comparing beta-based and read-based approaches, mixtures were also deconvolved using UXM \cite{loyfer2023dna}. This read-based deconvolution software was modeled on the same dataset by selecting the top 250 differentially methylated regions in one vs all other comparisons. After measuring the proportions of unmethylated, mixed, and methylated reads in the selected regions, NNLS optimization is used to infer cell type abundances in a composite sample. UXM infers percentages of 36 cell types, while my method is limited to 31 since I deemed some cell types inseparable, as previously explained. Given UXM utilizes a similar optimization algorithm, its main difference w.r.t. my proposed methods is the region selection process. 

\begin{figure}[ht!]
    \centering
    \includegraphics[width=0.9\textwidth]{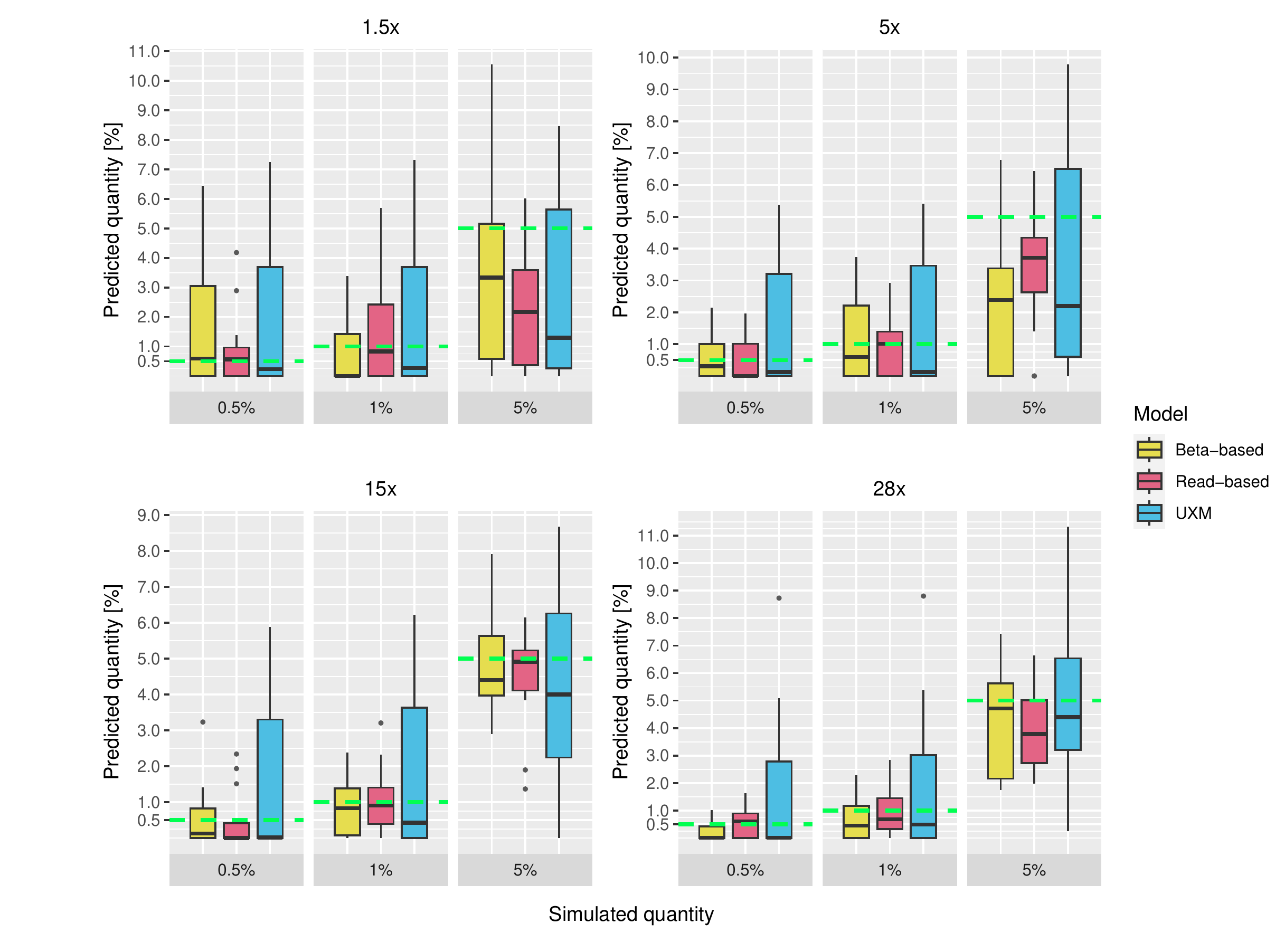}
    \caption{Predicted fraction of the minor element in the mixture across different sequencing coverage values. Green dotted lines denote the ground truth values.}
    \label{fig:predicted_true}
\end{figure}

\begin{figure}[ht!]
    \centering
    \hspace{4.2em}\includegraphics[width=0.75\textwidth]{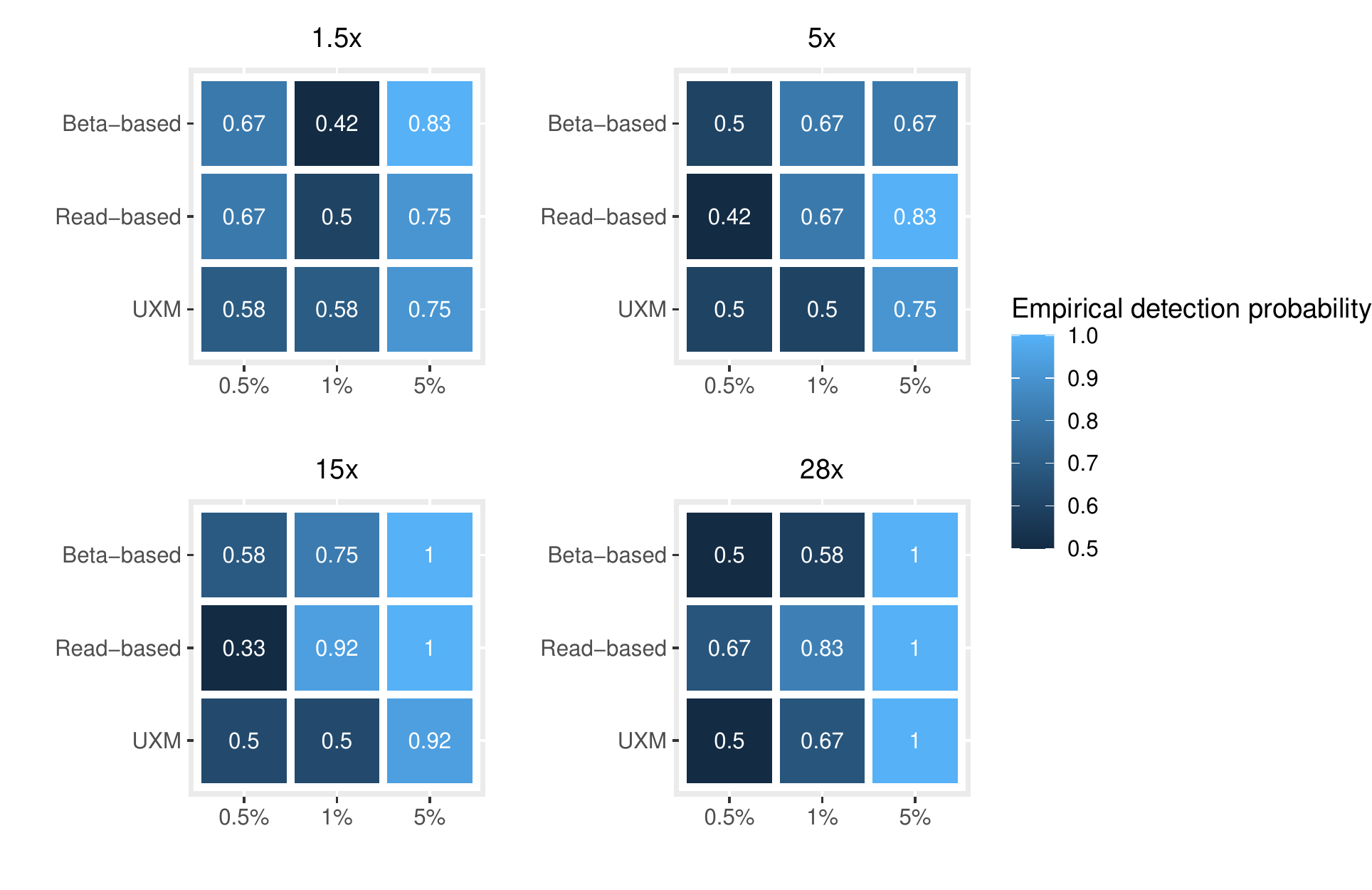}
    \caption{Empirical detection probability of minor cell type in mixture across different coverage values. A detection is counted if any non-zero percentage of the target cell type is reported.}
    \label{fig:dectection_prob}
\end{figure}

Figure \ref{fig:predicted_true} shows predicted data aggregated by minor cell type percentages and coverage. All methods struggle with detecting any non-zero fraction of the minor cell type at 0.5\% on all coverage values (Figure \ref{fig:dectection_prob}). With higher coverage and higher fractions, detection improves. No significant differences between tools were identified in terms of empirical detection probabilities. 

~\\

\begin{figure}[ht!]
    \hspace{-4.4em}
    \includegraphics[width=1.2\textwidth]{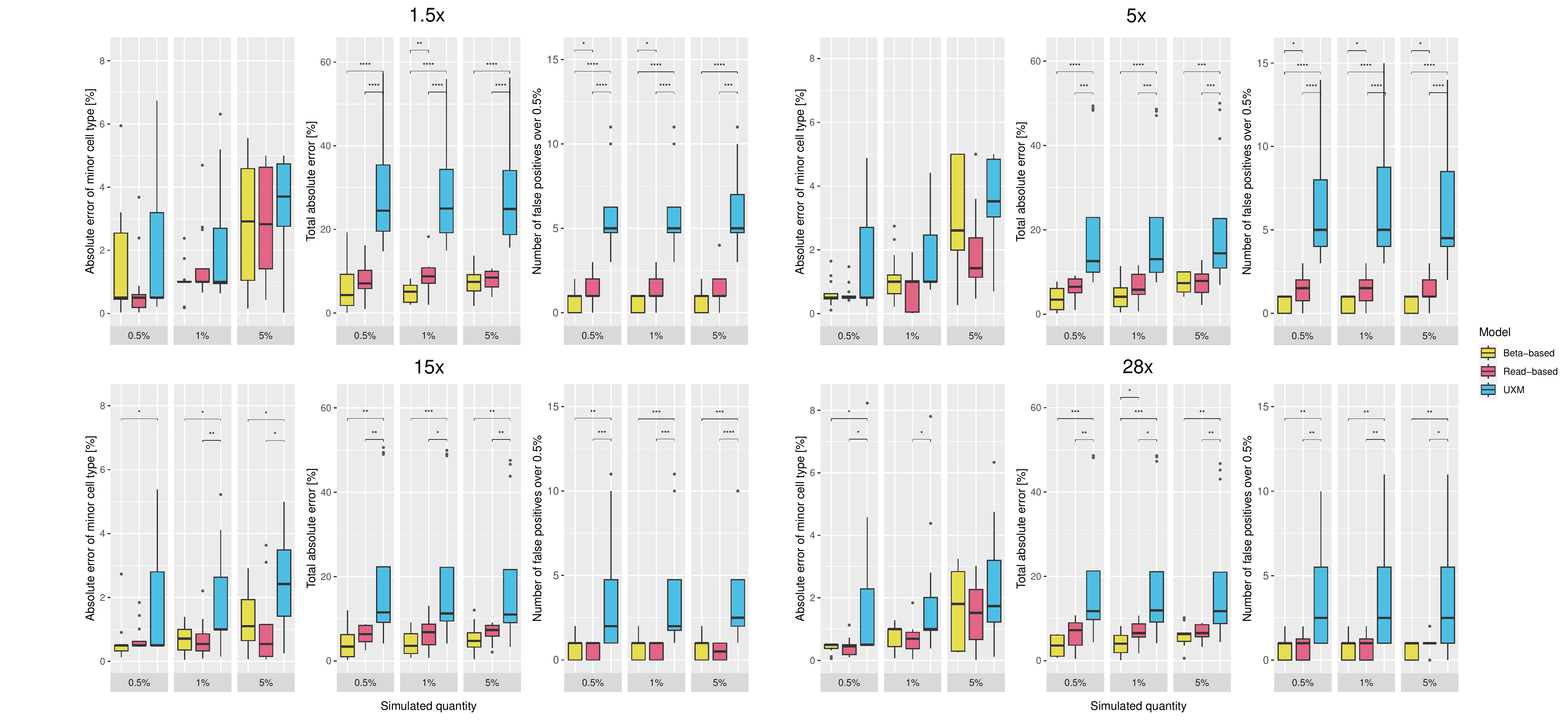}
    \caption{For each coverage value, absolute detection error of the minor element, total absolute error (based on Manhattan distance), and the number of false positives above 0.5\% is shown (left to right). Mann-Whitney U test was used for statistical significance.}
    \label{fig:error_merged}
\end{figure}

Next, deconvolution error was assessed by measuring the mismatch with the ground truth fraction of the minor cell type, total deconvolution error (expressed as Manhattan distance from the ground truth cell type vector), and by counting the number of false positives that appear in the mixture with quantities greater or equal to 0.5\%. These results are shown in Figure \ref{fig:error_merged}.

In terms of the absolute error of the minor cell type in the mixture, no significant differences are observed except in cases of 15x and 28x coverage simulations. However, total deconvolution error and the number of false positives above 0.5\% produced by any of my methods is significantly lower for all mixing rations and all coverage values w.r.t. UXM. 

\begin{figure}[ht!]
    \hspace{2em}\includegraphics[width=1\textwidth]{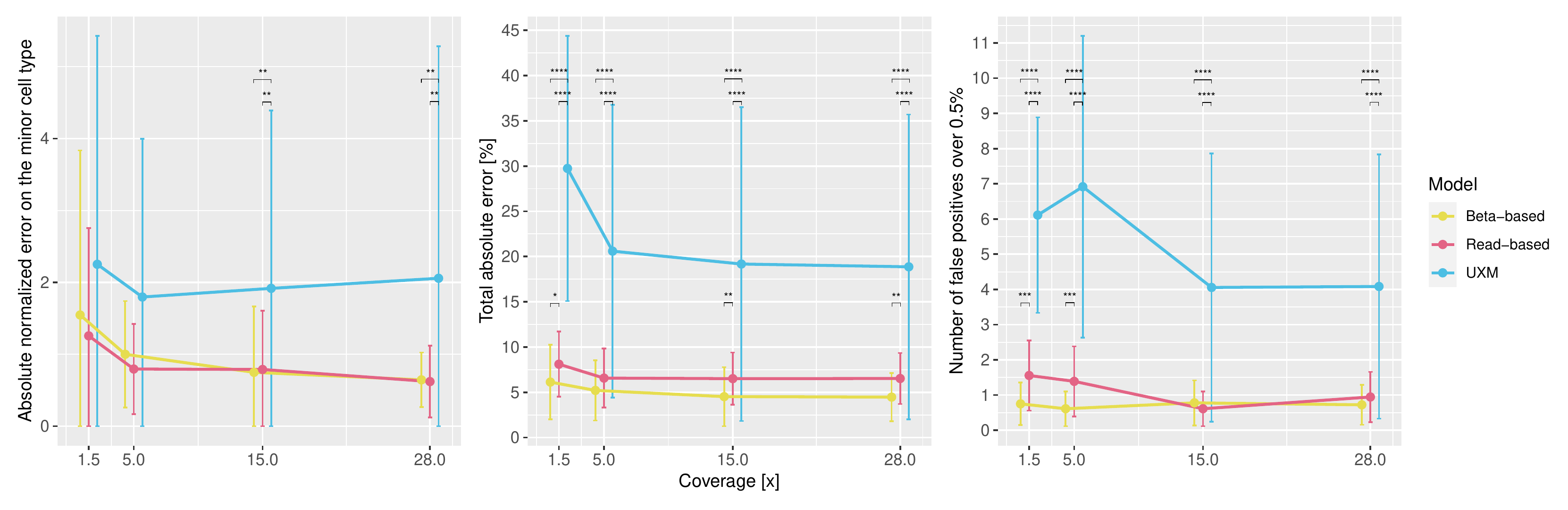}
    \caption{For each coverage value, the absolute detection error of the minor element, total absolute error (based on Manhattan distance), and the number of false positives above 0.5\% are shown (left to right). Mann-Whitney U test was used for statistical significance.}
    \label{fig:global_error}
\end{figure}

Figure \ref{fig:global_error} shows a more global result with more statistical power by aggregating all simulation results. To put equal weights on errors made when predicting different fractions of the minor cell type, errors were normalized by ground truth values being predicted for each sample. Here, the change of different types of error w.r.t. increasing sequencing coverage is obvious. Interestingly, even with 5x coverage, results are comparable with higher coverage values for all tools and all types of errors. Using my proposed methods, the normalized error of the minor cell type was significantly lower only in 15x and 28x experiments. However, in line with results shown in Figure \ref{fig:error_merged}, UXM demonstrated at least a 3-fold greater total error and a significantly higher number of false positives on all coverage values. Read-based and beta-based approaches produced similar results, with the exception of beta-based having less total error and false positives on some coverage values. 

In order to test whether introducing weights had any effect and to compare the proposed approach with UXM using the same optimization technique, the benchmark was run with the NNLS weighting turned off. Figure \ref{fig:uxm_global} shows UXM compared to unweighted beta and read-based deconvolution. The significance of improvements is almost identical to what is observed in Figure \ref{fig:global_error}. Figure \ref{fig:joined_rnb} shows the comparison between weighted and unweighted beta and read-based deconvolutions. Even though no significant differences were observed upon removing the weights, the mean error when using weighted NNLS is visibly lower in most cases. 

Simulating mixtures of different pairs of cell types at different ratios and sequencing coverage values allowed me to compare three models created on the same dataset. The proposed beta-based and read-based approaches, which utilize set cover feature selection, significantly outperformed a software based on the more common selection of regions via one-vs-all-other comparisons (UXM). Interestingly, in this benchmark, the beta-based method performed equally well and provided significantly less total deconvolution error and false positives w.r.t the read-based one. From these results, it can be concluded that for accurate deconvolution, the effect of the deconvolution level (read or beta value) is less important than how regions are selected.

\subsection{Deconvolution of leukocyte and composite tissue samples}

To further assess the proposed approach in different scenarios, I performed deconvolution on nine in-house peripheral blood mononucleated cells (PBMC) samples and six samples of different tissues obtained from the Roadmap database. As the beta-based method was deemed the most accurate in the benchmark, it was the only one used. Additionally, unlike the read-based method, it does not require sequence data in BAM format -- increasing its applicability, as sequence data is often hard to obtain and store. Based on the previous sensitivity analysis, in these experiments, the detection threshold was set to 0.45\%, as this value is just below the mixing fraction deemed the detection limit. All cell types detected below this threshold are removed, and the output is re-normalized to sum to 1. 

\begin{figure}[ht!]
    \centering
    \hspace{2em}\includegraphics[width=0.9\textwidth]{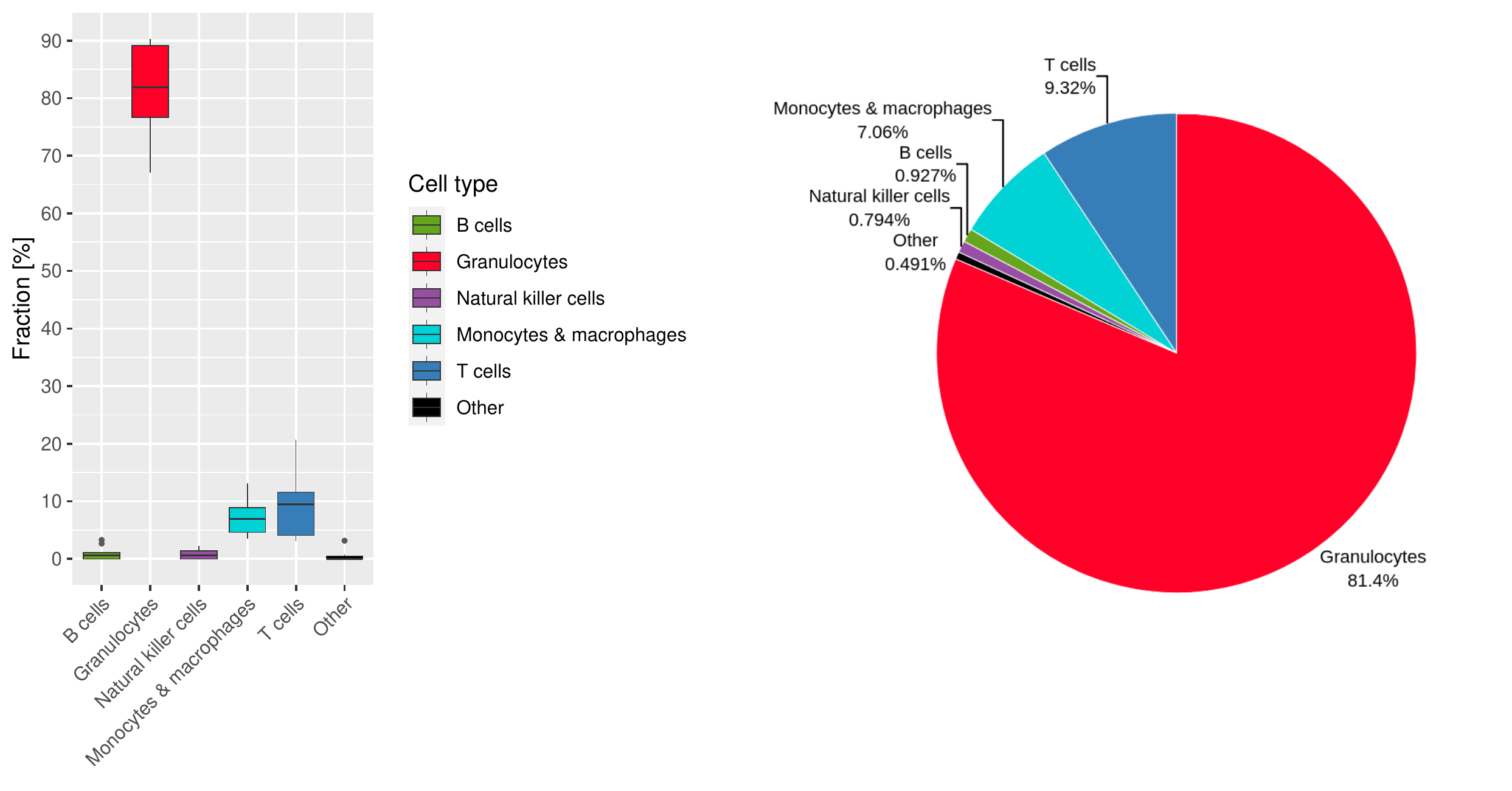}
    \caption{PBMC decomposition of nine samples (left); mean cell type values normalized to sum to 1 (right).}
    \label{fig:leukocytes}
\end{figure}

~\\
Figure \ref{fig:leukocytes} shows the composition of blood cells in nine PBMCs which were used as germline pairs in a cancer study \cite{beltran2020circulating}. All samples are primarily comprised of granulocytes with smaller fractions of T cells and monocytes/macrophages. These results suggest a higher fraction of granulocytes than reported in previous decompositions of such samples \cite{loyfer2023dna}.
The label "Other" (black) summarizes all cell types which are impossible to find in a given sample -- e.g. finding prostate or breast epithelia in a PBMC sample. These can be thought of as false positives. 

~\\

\begin{figure}[ht!]
    \centering
    \hspace{2em}\includegraphics[width=1\textwidth]{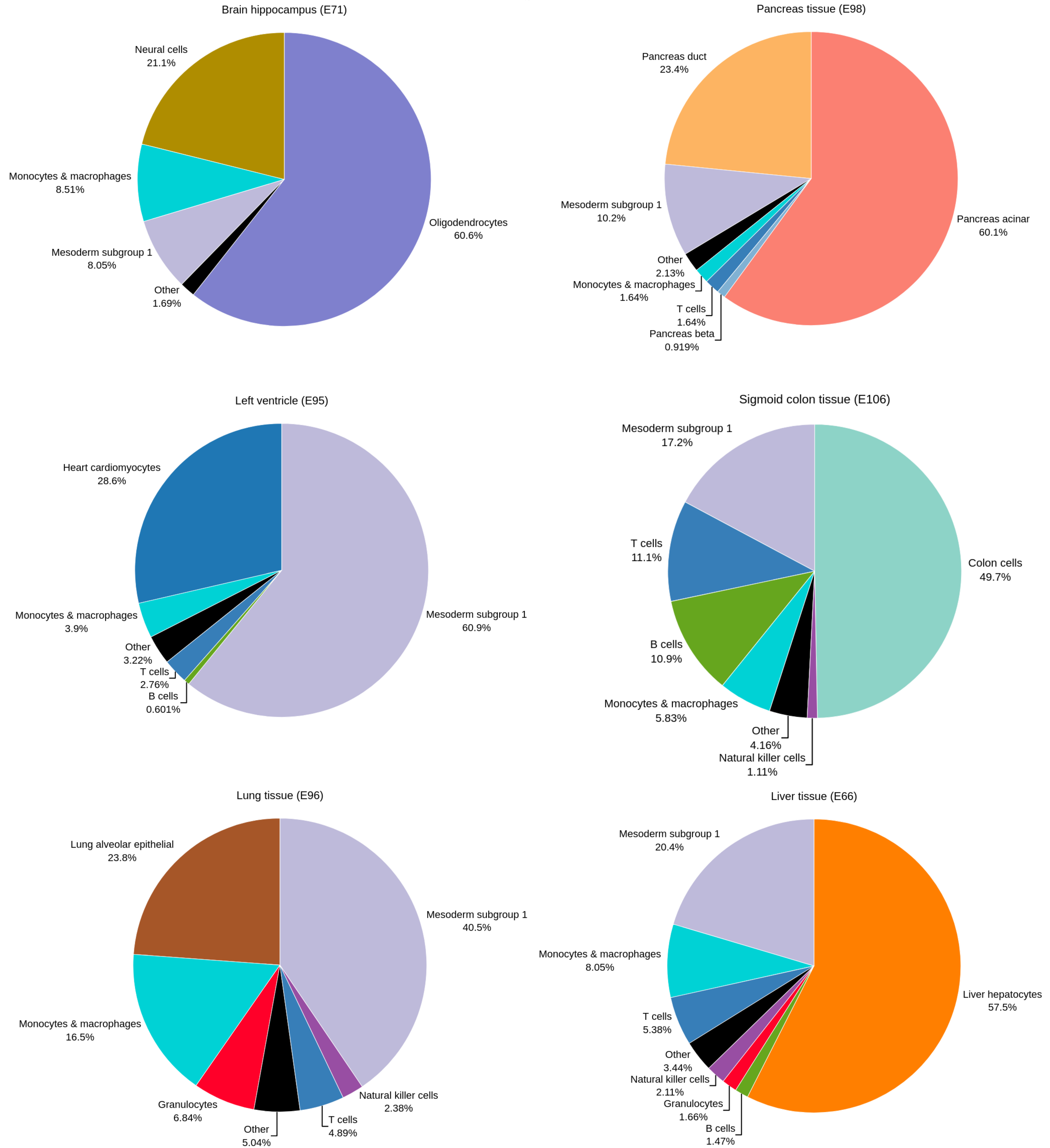}
    \caption{Inferred cell type fractions of Roadmap tissue samples.}
    \label{fig:tissue}
\end{figure}

Next, six WGBS samples originating from different tissues were retrieved from the Roadmap project, and their purity was assessed. In all samples, the primary cell types of a given tissue were identified at a high fraction (Figure \ref{fig:tissue}). Alongside these, I observed a high ratio of mesoderm subgroup 1 (joined group of endothelial, smooth muscle, skeletal muscle, and adipocyte cell types which I previously assessed as inseparable) and immune cells - especially monocytes and macrophages (8.9\% on average). A high fraction of oligodendrocytes and neural cells were identified in brain hippocampus tissue. This sample also had the lowest fraction of endothelial/muscle cell types. Pancreas tissue was assessed to be composed mostly of acinar (60\%), followed by duct cells with the lowest fraction of monocytes and macrophages. Colon tissue consisted primarily of colon epithelial and endocrine cells. This sample also contains the highest fraction of B cells (10.9\%). Surprisingly, both ventricular and lung tissue samples had fractions of cardiomyocytes and lung alveolar epithelia less than 30\%, respectively. These are primarily composed of mesoderm subgroup 1 cell types. Moreover, a low fraction of alveolar epithelia was also observed by \textit{Loyfer et al}. in the same sample using UXM. Again, known false positives were joined under the "Other" label.

%% file: chapters/Methods/methods.tex
\subsubsection{WGBS cell type data retrieval and pairwise DMR 
identification}

Data generated by \textit{Loyfer et al.} consisting out of 186 samples from 44 cell types, mapped to hg19, was retrieved from NCBI GEO database (GSE186458) in both pat format, which contains read-level information and beta format, which contains beta values for all 28.22 million CpG sites. Beta files were converted to tables using wgbstools beta\_to\_table module \cite{loyfer2023dna}. These tables were then used for coverage analysis and DMR identification.
For a list of all samples used in this research project, refer to the \textit{Data availability} section.

For all 946 pairwise comparisons of 44 cell types, Metilene (version 0.2-8) \cite{juhling2016metilene} was used for calling DMRs. The minimal number of CpG sites in a DMR was set to 5, and Benjamini-Hochberg correction was used (-m 5-c 2 parameters). To filter raw Metilene output, the metilene\_output.pl module was used with the maximal FDR value of 0.05 and minimal difference in mean region methylation of 0.4 (-p 0.05 -d 0.4 parameters). As for some pairs of biologically related cell types, the number of DMRs was low, these were joined, which reduced the number of cell types from 44 to 38. Pairwise DMR calling procedure was re-applied for all combinations of these 38 cell types. This resulted in 703 bed files containing high-quality DMR calls between all cell types. 

\subsubsection{Extraction of UMRs}

To identify sets of UMRs for each cell type, all 703 pairwise DMR identification files were intersected using bedtools multiinter (version 2.27.1) function \cite{quinlan2010bedtools}. Then, seed markers were identified for all cell types. Seed markers of a given cell type are regions that are identified as differentially methylated in all pairwise comparisons that include that cell type. These markers were then extended downstream and upstream to include regions supported by at least 97\% of pairwise comparisons that include the considered cell type identified as differentially methylated. The threshold of 97\% means no more than one, out of 37 pairwise comparisons that include a given cell type, can omit to call a DMR for a region for it to be used as a seed extension. This approach was conducted separately for hypo and hypermethylated markers across all cell types. For all putative markers, mean beta values and mean coverage was calculated from all cell types' data. Only markers with at least 3 CpG sites and mean coverage of 15 were considered. Bed files containing region information, UMR type, methylation values, and CpG positions were created for all cell types (See \textit{Data availability} chapter). As a sanity check, UMR sets were checked for overlaps between cell types, and none were found. When assessing the number of UMRs also identified by \textit{Loyfer et al.}, an intersection of 10\% was used (bedtools intersect -f 0.1). 

\subsubsection{Functional element analysis}

To assess the distribution of functional elements in hypomethylated and hypermethylated UMRs, various gene annotations for hg19 reference were retrieved using the UCSC table browser. Annotations for 3' and 5' UTRs, exons, introns, promoters (1 kb upstream of the TSS), and downstream regions (500 bp downstream of 3' UTR) were combined with databases of 1988217 enhancers \cite{gao2020enhanceratlas} and 8827 silencers \cite{zeng2021silencerdb}. Only experimentally validated elements were considered in both resources. To count an overlap with a functional element, a UMR has to share at least 40\% of its length with it (bedtools intersect -f 0.4). Therefore, a single element can fall into multiple functional elements. If a UMR does not overlap any functional annotation with this criteria, it is counted as "Unknown". Fisher's exact test was used to test the differences in ratios of different elements.

\subsubsection{Gene enrichment analysis}

Gene enrichment analysis was performed for all cell types independently by separating hypo and hypermethylated UMRs using rGREAT (version 2.0.2) \cite{gu2023rgreat}. This tool envelopes an R-based API for the GREAT \cite{mclean2010great} web-based enrichment analysis software through the submitGreatJob function. In all rGREAT analyses, only enrichment for Gene Ontology terms was performed using the oneClosest gene rule and hg19 as reference. As the retrieved data is unfiltered, only statistically significant terms were considered where both Benjamini-Hochberg corrected binomial and hypergeometric p-values were less than 0.05. In the results, only corrected hypergeometric p-values are presented. The regions associated with each significantly enriched term and their closest genes with TSS distances were retrieved using the getRegionGeneAssociations function. 

\subsubsection{Motif enrichment analysis}

Motif enrichment was run separately for hypo and hypermethylated UMRs on individual cell types and by piling up UMRs of the same type. HOMER's findMotifsGenome.pl (version 4.11.1) \cite{heinz2010simple} was used with --size 250 and -bits arguments, and by specifying hg19 as a reference. To obtain regions with binding sites of interest and their closest genes, the annotatePeaks.pl function was used. Motifs of interest were provided via the -m parameter. As HOMER does not perform p-value adjustment, only highly significant results in the known category were considered with a p-value less than 1-e8. 

\subsubsection{Assessing chromatin states in hypomethylated UMRs}

Chromatin state annotations from the 15-state ChromHMM model were retrieved from Roadmap (\rurl{egg2.wustl.edu/roadmap/data/byFileType/chromhmmSegmentations/ChmmModels/coreMarks/jointModel/final}). Bed files annotated as dense were downloaded for samples E27 (breast myoepithelium), E29 (monocyte), E32 (b cell), E30 (neutrophil), E46 (natural killer cell) and E66 (liver tissue). These were compared with UMRs of both corresponding and non-corresponding cells by observing overlaps. In this case, a UMR is assigned to a ChromHMM annotation if there is at least 90\% overlap (bedtools intersect -f 0.9). UMRs that do not overlap any annotations with this criteria are assigned with the "Unmatched" label. To test the significance of annotation enrichment in matched vs non-matched UMR-annotation pairs, Fisher's exact test was used.

\subsubsection{Deconvolution pipeline}

\noindent\textbf{\small 1. Genome segmentation and region filtering for deconvolution}
\vspace{0.75em}

\noindent To segment the whole genome into methylation-informative regions, all 946 pairwise DMR bed files for all cell types were intersected using bedtools multiiter module. This resulted in 9.1 million regions that had to be filtered. For all regions covering more than three CpG sites and having a density of at least 2.5 CpGs/100bp, mean beta value, mean coverage, and intra-cell-type standard deviation were calculated for each cell type from its samples. Regions where all cell types have coverage greater than 17.5 and a standard deviation less than 0.33 were retained.

The second filtering step included selecting only regions where cell types group into two distinct partitions, $p1$ and $p2$, separated by at least 0.4 delta-beta. Let $T=\{t_1,t_2,...,t_{N}\}$ be a set of all cell types, where  $N=44$, and let $q_i=(q_1^i, q_2^i,...,q_N^i)=(q_a^i)_{a=1}^N$ be a vector of mean cell type beta values in a region with index $i$, sorted in ascending order. Region with index $i$ was retained only if there exists $a\in\{1,2,...,N-1\}$ such that $q^i_{a+1} - q^i_{a} \geq 0.4$. This filtering allows one to classify each cell type as $p1$ or $p2$ in each region. Additionally, in regions where $q^i_{a+1} - q^i_{a} \geq 0.4$ for some $a\in\{1,2,...,N-1\}$, intra-partition delta-beta (distance between partitions) was defined as $b_i = q_{a+1} 
- q_{a}$.
After these filtering procedures, $L_f = 28477$ regions were retained.

\vspace{1em}
\noindent\textbf{\small 2. Aggregating inseparable cell types}
\vspace{0.75em}

\noindent After filtering, it was necessary to assess which cell-type pairs have enough discriminative regions to be reliably separated and which ones must be joined.

 For the set of cell types $T$, there are $M=946$ possible pairwise combinations; therefore, there are M possible cell-type separators. An ordered set of all possible separators $ Z=(Z_{1}=\{t_1, t_2\},Z_{2}=\{t_1,t_3\},...,Z_{M}=\{t_{43}, t_{44}\})$ can be created, thus enumerating them. Given a filtered region index $i$, let $P^i_1$ and $P^i_2$ be non-overlapping sets of cell types that fall into $p1$ and $p2$ partitions of $i^{th}$ filtered region, separated by at least 0.4 delta-beta. The filtered region with index $i$ is considered to be able to separate a set pair of cell types $\{t_a$,$t_b\}$, where $a,b \in \{1,2,...,N\} \wedge a \neq b$, if $(t_a \in P^i_1 \wedge t_b \in P^i_2) \vee (t_b \in P^i_1 \wedge t_a \in P^i_2)$ (referred to as \textbf{ $\star$} criteria). For $i^{th}$ filtered region, its set of separators can be denoted as $C_i = \{C_1^i, C_2^i,...,C_{m_i}^i\}=\{C^i_{j}\}_{j=1}^{m_i}$, where $m_i$ is the total number of separators for that region and $C_j^i=\{t_{a}, t_{b}\}$ is a set pair of separable cell types that satisfies (${\star}$) for any $C_j^i \in C_i$. This set of separators is computed as a set of all possible pairwise combinations of cell types in the two partitions: $C_{i}=\{ \{t_a,t_b\}:
t_a \in P^i_1 \wedge t_b \in P^i_2\}$.
By using $Z$ as an indexing reference, a vector $v_i$ can be created denoting whether $i^{th}$ filtered region has a given separator or not:
~\\
\vspace{0.5em}
\hspace{4em}
$v_i = (v_1^i, v_2^i, ..., v_M^i) = (v_k^i)_{k=1}^{M}$  such that $v_k^i =$
$
\begin{cases}
    1,              & \text{if } C_j^i = Z_{k} \text{ for any } C_j^i \in C_i\\
    0,              & \text{otherwise}
\end{cases}
$

\vspace{0em}

By performing separator indexing on all regions, it was possible to obtain the total number of each separator in the filtered region set: $v_{global} = {\sum_{i=1}^{L_f} v_i}$, where $L_f=28477$ is the size of the filtered region set. As values of $v_{global}$ represent the total number of separators between each pair of cell types, creating a symmetric, 0-diagonal $(N \times N)$ similarity matrix $J$ is possible. $J$ was used to construct an adjacency matrix $A$ by thresholding in the following way:

\vspace{0.5em}
\hspace{6em}
$A=(A_{ab})_{a=1,2,...,N; b=1,2,...,N}$ such that $A_{ab}=$
$
\begin{cases}
    1,& \text{if } J_{ab} < 400 \\
    0,& \text{otherwise}
\end{cases}
$
\vspace{0.7em}

The threshold of 400 was selected because it provided a good tradeoff between generating as many final region subsets as possible and separating many important cell types. 
Using a set $E=\{ \{t_a, t_b\}: A_{ab} = 1 \text{ for } a,b \in \{1,2,...,N\}\}$
an undirected graph $G(N, E)$ was created. Each of the five connected components in the set of connected components $C(G(N, E))$ was aggregated into a single group. 1) CD4 and CD8 T cells; 2) monocytes and macrophages; 3) colon epithelia and endocrine; 4) kidney podocytes and kidney tubular epithelia; and 5) adipocytes, all endothelia, smooth and skeletal muscles were joined. The proximity between merged cell types aligns with DMR distances discussed in Figure \ref{fig:dmr_distance_heatmap}. All regions in the filtered region set that had separators between members of the connected components were excluded from the region pool to reduce variability in aggregated cell types.
This joining procedure resulted in a set of $L_c=26476$ candidate regions and a reduction from $N=44$ to $N_s=31$ cell types which were inferred in deconvolution.

\vspace{1em}
\noindent\textbf{\small 3. Region selection via set cover approximation}
\vspace{0.75em}

\noindent Let $V$ be an $(M \times L_c)$ matrix where each row represents the presence of $Z$-indexed separators in $Lc$ candidate region. Let ${I} = (I_i)_{i=1}^{L_c}$ be an ordered set, where $I_i = \{k: V_{ik} = 1 \text{ for } k=1,2,...,M\}$ is a $Z$-based, separator index set of the $i^{th}$ candidate region; and let $S$ be a $(L_c \times N)$ matrix of intra-cell-type standard deviation of beta values of each region. Additionally, let $rescale(X): \mathbb{R}^n\to [a,b]^n:$ be a function for rescaling an $n$-dimensional vector into a bounded [$a,b $] interval.
$$rescale(X) = \frac{(b-a)(X - \min(X))}{\max(X) - \min(X)} + a$$

Then, ${\widetilde{b}} = rescale\big((b_i)_{i=1}^{L_c}\big)$ can be defined as a vector of inter-partition delta-betas, rescaled to [$1,2$] interval; and $s^{b}=\smash{\displaystyle\max_{i=1,2,...,L_c}}S_{ia}$ can be defined a vector of maximal intra-cell-type standard deviations across cell types. ${\widetilde{s^b}=rescale(s^b)}$ represents this vector scaled to [$1, 2$] interval. 

The goal of the following region selection was choosing a close-to-minimal subset of regions that covers the set of separator indices ${X} = \{1,2,...M\} - R$, where $R = \{k: Z_k \in E\}$ corresponds to $Z$-based indices of separators of aggregated cell types. This procedure is described in Algorithm \ref{alg:set_cover}.

\RestyleAlgo{ruled}
\SetAlCapNameFnt{\small}
\SetAlCapFnt{\small}

\begin{algorithm} [!ht]
\caption{Greedy weighted set cover (${V}$, ${I}$, $\widetilde{b}$, $\widetilde{{s^b}}$, ${X}$, $M$, $Selected = \emptyset$)}
\label{alg:set_cover}
$U \gets {X}$ ~\\
$Q \gets \emptyset$ ~\\
$W \gets \overrightarrow{0}_{M}$ ~\\
\While{$U \neq \emptyset$}{
    
select region index $i:i \notin Selected$ that maximizes
$\dfrac{|I_i \cap U| \widetilde{b}_i} 
{\Big(\sum_{k=1}^{M} (V_{ik} W_k)+1\Big)\widetilde{s^{b}_i}}$

\If{$I_i \cap U=\emptyset$} 
{
    \textbf{return} $\emptyset$
     \text{\footnotesize ~~~~ $\triangleleft$ case in which separators cannot be covered}
} ~\\

$U \gets U - I_i$ ~\\
$Q \gets Q \cup i$ ~\\
$W \gets W + W_i$
}
\textbf{return} $Q$
\end{algorithm}

A single run of Algorithm \ref{alg:set_cover} returned indices of regions that form a subset that can cover $X$, as one shown in Figure \ref{fig:single_basis}. In the maximizing term in line \textbf{5}, the score of each region is weighted proportionally to the inter-partition delta-beta value and inversely proportionally to the cell-type-wise maximal standard deviation of that region. The first term in the denominator forces the algorithm to prefer regions with fewer covered separators, thus maintaining uniformity in the number of separators.

To mitigate the biological variability of methylation, a single region subset is insufficient for deconvolution. Algorithm \ref{alg:iterative} shows the iterative selection of region subsets until depletion of any separator. The maximal number of possible iterations was 369, resulting in  $L=10479$ selected regions in total. 

\begin{algorithm} [!ht]
\caption{Iterative region selection (${V}$, ${I}$, $\widetilde{b}$, $\widetilde{s^b}$, ${X}$, $M$)}
\label{alg:iterative}
$Selected \gets {\emptyset}$ ~\\

\While{$1$}{
$Q \gets$ {Greedy weighted set cover (${V}$, ${I}$, $\widetilde{b}$, ${\widetilde{s^b}}$, ${X}$, $M$, $Selected$)} \\
    \If{ $ Q = \emptyset $} 
    {
        \textbf{return} $Selected$
    } ~\\
$Selected \gets Selected \cup Q$
}
\end{algorithm}

\vspace{1em}
\noindent\textbf{\small 4. Creation of deconvolution atlas}
\vspace{0.75em}

\noindent For the selected regions, atlas reference values were created separately for the beta-based and read-based methods. Considering separable $N_s=31$ cell types (including aggregated groups) and $L$ selected regions, the beta-based atlas was defined as a ($L \times N_s$) matrix $A^b$ containing mean region beta values of all cell types across all selected regions. Values of aggregated cell types were computed as the mean of their members.

As for the read-based atlas, the fractions of reads in unmethylated partitions $p1$ of each region were calculated. Given a selected region index $i$, let $Q_i^1$ and $Q_i^2$ be sets of cell types beta values in $p1$ and $p2$ partitions of the $i^{th}$ selected region. As defined in the filtering procedure, $\min (Q_i^2) - \max{(Q_i^1)} \geq 0.4$ for any $i$ in the selected region set. Let $g = (g_i)_{i=1}^{L}$ be a vector of region-specific thresholds, where $g_i = \max{(Q_i^1)} - \dfrac{\min{(Q_i^2)}}{2}$, which corresponds to the midpoint between the inner boundary beta values of partitions of the $i^{th}$ selected region. In each sample independently, reads with more than three CpG sites that overlap a selected region are classified based on the fraction of methylated CpG sites. Given a read $r$, in $i^{th}$ region with $n$ total CpG sites and $m$ methylated sites, if $\dfrac{m}{n}$ $\leq g_i$, $r$ is classified as $p1$, otherwise as $p2$. After all reads have been classified in this way, the fraction of $p1$ reads is determined independently for each selected region.
For all 186 samples covering all $N$ cell types, for all selected regions $i=1,2,...,L$, reads overlapping three or more CpG sites were classified based on the region-specific thresholds $g_i$ and the fraction of $p1$-class reads was calculated. Reference atlas values of cell types were taken as the mean $p1$-fraction of cell type samples, while fractions of aggregated cell types were calculated as means of group members - resulting in an $(L \times N_s)$ matrix $A^r$. Additionally, as intra-cell-type standard deviation of $p1$ fractions was computed for all $N$ cell types, $s_r$ represents an $(L \times 1)$ vector of maximal cell-type-wise standard deviations of $p1$ fractions. Fractions of reads were calculated from pat files using wgbstools.

\vspace{1em}
\noindent\textbf{\small 5. Fitting cell proportions}
\vspace{0.75em}

\noindent Starting from a Bismark-mapped \cite{krueger2011bismark} BAM file, the reads overlapping selected regions are extracted using bedtools intersect. Then, using samtools \cite{danecek2021twelve} (version 1.15.1), these reads are filtered, leaving only primary alignments with MAPQ values of at least 20. For read-based deconvolution, in all selected regions, fractions of methylated CpG sites on reads overlapping at least three region CpGs, were compared to the region-specific threshold $g$, and those reads in partition $p1$ were counted. If possible, whole read pairs were assessed instead of reads for data generated in paired-end sequencing experiments. If both pairs were in the same region, the number of CpG sites covered by the whole fragment had to be at least three for it to be considered. Counting $p1$-classified reads resulted in an $(L \times 1)$ vector $y^r$. Additionally, let $C$ be an $(L \times L)$ matrix where the diagonal elements correspond to the total number of reads in both partitions ($p1$ and $p2$) and all others are 0; and let $\widetilde{s^r_{inv}} = rescale(\overrightarrow{1}_{L} - s^r)$ be a vector of inverted maximal $p1$-fraction standard deviations, rescaled to [$1,2$] interval. Obtaining the $(N_s \times 1)$ vector of cell type abundances $t$ was performed by minimizing $|\widetilde{s^r_{inv}} \cdot (C \times A^r \times t - y^r) |_2$, subject to non-negative $t$, where $A^r$ is the $p1$-fraction atlas.

Preparation for beta-based deconvolution involved the same procedure of BAM processing and read filtering. Instead of counting $p1$-classified reads, beta values of selected regions are computed using wgbstools, resulting in a $(L \times 1)$ vector, $y^b$. Let $\widetilde{s^b_{inv}} = rescale(\overrightarrow{1}_{L} - s^b)$ be a vector of inverted maximal beta value standard deviations, rescaled to [$1,2$] interval.
Obtaining $t$ was performed by minimizing $|\widetilde{s^b_{inv}} \cdot (A^b \times t - y^b) |_2$, subject to non-negative $t$, where $A^b$ is the beta value atlas.

In both read and beta-based deconvolution, the output of weighted NNLS is normalized, so its sums to 1. Additionally, optimization was performed with multi-starting -- running optimization several times from random initial values and then choosing the result with the lowest L2 distance. Scipy version 1.10.0 was used for optimization. Both types of deconvolution are implemented in a tool presented in this research project called methylcover (See \textit{Data availability} chapter). Apart from beta-based and read-based deconvolution, methylcover also has a prepare\_bam module which envelopes the pipeline for processing BAM files into a form suitable for deconvolution.  

\subsubsection{Synthetic mixture simulations}
Four WGBS, sorted-cell array samples were retrieved from the Blueprint database \cite{fernandez2016blueprint} in hg38-mapped files. These included natural killer cells (EGAF00000733380), CD8 T cells (EGAF00000521928), monocyte (EGAF00000498309), and B cells samples (EGAF00000733412). All BAM files were converted into pairs of FASTQ files (samtools bam2fq) and were realigned with Bismark (bismark -q -X 500) to hg19. Hg19-mapped BAM files were then deduplicated (samtools fixmate -m; markdup -r). Synthetic mixtures contained only two cell types in known quantities, one as the minor part, the other as the background. These were mixed in rations of 95/5\%, 99/1\%, and 99.5/0.5\, on 1.5x, 5x, 15x and 28x mean sequencing coverage values. Combining all pairwise sample combinations in all mixing ratios on all coverage values resulted in 144 samples. To create mixtures, samtools view was used with the random subsampling option -s. The background sample is downsampled to have the desired coverage value; then, based on the number of reads, a fraction of the minor cell type is added. 
The resulting BAM files were converted to pat format to deconvolve with UXM, and the prepare\_bam module was used to produce the right input format for beta and read-based deconvolution. While benchmarking, methylcover was used with a detection threshold of 0 (-dth argument)  and UXM was used with default parameters.

\subsubsection{Roadmap and PBMC sample deconvolution}

A set of nine PBMC samples was retrieved from the study by \textit{Beltran et al.} \cite{beltran2020circulating} in BAM format. Input files for beta-based deconvolution were produced with methylcover prepare\_bam functionality. Deconvolution was run with the detection threshold set to 0.45\% (-dth 0.0045), meaning that any cell type detected below 0.45\% is erased from the final result, and the output is renormalized to sum to 1. 

Data for six Roadmap samples (E66, E71, E95, E96, E98 and E106) was retrieved from the Roadmap database (\rurl{egg2.wustl.edu/roadmap/data/byDataType/dnamethylation/WGBS}) in bigWig format for coverage and fractional methylation. Using UCSC's utility bigWigAverageOverBed, coverage and beta value formats were processed and combined to produce input files for beta-based deconvolution. The detection threshold was set to 0.45\%.

\subsubsection{Statistical analysis and visualization}
All statistical tests were conducted using a two-sided Mann-Whitney U test. For functional element distribution analysis and ChromHMM annotation enrichment, Fisher's exact test was used. The number of "*" symbols in figures denotes the level of statistical significance: * - p-value $\leq 0.05$; ** - p-value $\leq 0.01$; *** - p-value $\leq 0.001$; **** - p-value $\leq 0.0001$. All visualizations were produced in R using ggplot2 in combination with ggpubr, except cell-type composition pie charts for which the plotly package was used.

%% file: chapters/Conclusions/conclusions.tex
In this research project, I introduced a method for identifying unique methylation signatures of 44 healthy cell types, investigated their functional properties, and created a novel deconvolution approach where discriminative regions are selected via set cover algorithm based on pairwise comparisons.

Performing pairwise differential methylation analysis revealed that methylomes group in ways that reflect their developmental origins and biological functions, in line with previous findings \cite{loyfer2023dna}. However, there are exceptions to this trend, as physiologically different cell types, like adipocytes and endothelia, can converge to having very related methylation profiles. Combining regions resulting from pairwise differential methylation allowed me to identify a set of unique methylation markers for each cell type. Since some pairs of cell types had a high degree of similarity, they were joined before markers were extracted. This reduced the number of cell types to 38 but also increased the possibility of capturing markers in aggregated cell types. All the cell types chosen for aggregation shared high functional similarity, except smooth muscle and fibroblast. Identified markers consist mostly of uniquely demethylated regions ($\sim$98.5\%), and their total number proved to be variable across cell types. Uniquely methylated markers are rarer but significantly longer, richer and denser in terms of CpG sites, which is explained by the fact that more than half of them are located in CpG islands. Cell-type-specificity of these markers was verified using gene and motif enrichment analysis. Besides validating UMRs, motif enrichment uncovered many potential novel transcription factor-gene interactions in hypomethylated UMRs. However, to understand these interactions better, further research is needed.  

Regarding functional elements, markers are most frequently found in enhancer and intronic regions. As the effect of DNA methylation is versatile in terms of increasing or decreasing gene activity and transcription factor binding, a future perspective is inferring transcriptional states of genes using the methylation patterns surrounding them. For this purpose, an integration with a large body of transcriptional cell-type-specific data is needed, possibly leveraging single-cell data. By using ChromHMM chromatin annotations, I inferred that most hypomethylated UMRs are in enhancer-like regions where the only histone modification is H3K4me1. This enrichment aligns with functional element distribution results and was only observable when annotations for corresponding cell types were used, thus validating cell-type-specificity with orthogonal data. 

As the number of UMRs is not uniform across cell types, and can be as low as a single one, another method for selecting regions for the purpose of deconvolution was needed. Using the sets of pairwise DMRs identified in previous analyses, I implemented a novel method for discriminative feature selection using the set cover algorithm. To my knowledge, this is the first application of such an algorithm for genomic feature selection. Both beta-based and read-based deconvolution methods were developed using a set of regions selected in such a manner. As no publications have compared the performance of these two methods on the same region set, a rigorous comparison was an additional goal of this research project. Beta and read-based methods were compared on simulated mixtures alongside UXM, a deconvolution tool built on the same reference data. Overall, both methods that used set cover region selection exhibited at least a 3-fold reduction in error compared to UXM, and this difference was mainly dictated by a more precise identification of the background cell-type component in deconvolution. Surprisingly, even though the beta-based approach is less preferred than read-level deconvolution in recent literature, in this benchmark, it did not only perform as well as the read-based approach, but also at times was more robust to false positives. This suggests that the level at which deconvolution is performed is less important than the process of selecting informative regions.  


Using the set of selected regions, if the experiment aims to identify the fractions of referenced cell types, one might design a custom, targeted sequencing assay, thereby significantly reducing the experimental cost, improving coverage, and leading to easier data processing and handling. As all selected deconvolution regions were used in this project, further research aims to identify the minimal number of set cover region subsets capable of performing accurate deconvolution. Additionally, the proposed feature selection applies not only to methylation but, in theory, to any multi-class high-dimensional data problem where the goal is to find minimal subsets able to separate cell types. In general, identified UMRs and deconvolution regions are proven to be valid for healthy adult cell types. How age and pathologies known to cause destabilization of the epigenome might affect these methylation patterns is still unknown, therefore expressing the need to expand the reference dataset with samples of various conditions.
In the case of genome-wide methylation disruptions occurring in cancer, it is possible that many regions selected for deconvolution are affected, thus confounding the inferred cell type abundance results. Another future research goal is to expand the current set of cell types with pan-cancer data. A deconvolution tool built on such a reference set could have numerous clinical applications such as non-invasive cfDNA cancer detection and tumor purity estimation. 

%% file: chapters/Data/data.tex
The deconvolution tool created in this project is called methylcover and is available at \rurl{github.com/ciganche/methylcover}. It infers the relative abundance of 31 cell types and supports hg19 and hg38 human genome references. It is comprised of modules for beta and read-level deconvolution, as well as a utility for preparing input formats from BAM files. 

The set of UMRs, motif and enrichment data, as well as markers with specific binding sites, are available at \rurl{bit.ly/3IJMtYE}.

%% file: chapters/Supplementary/supplementary.tex
\setcounter{equation}{0}
\setcounter{figure}{0}
\setcounter{table}{0}
\renewcommand{\theequation}{S\arabic{equation}}
\renewcommand{\thefigure}{S\arabic{figure}}
\renewcommand{\thetable}{S\arabic{table}}

\begin{figure}[ht!]
    \centering
    \hspace{-1.3em}
    \includegraphics[width=0.93\textwidth]{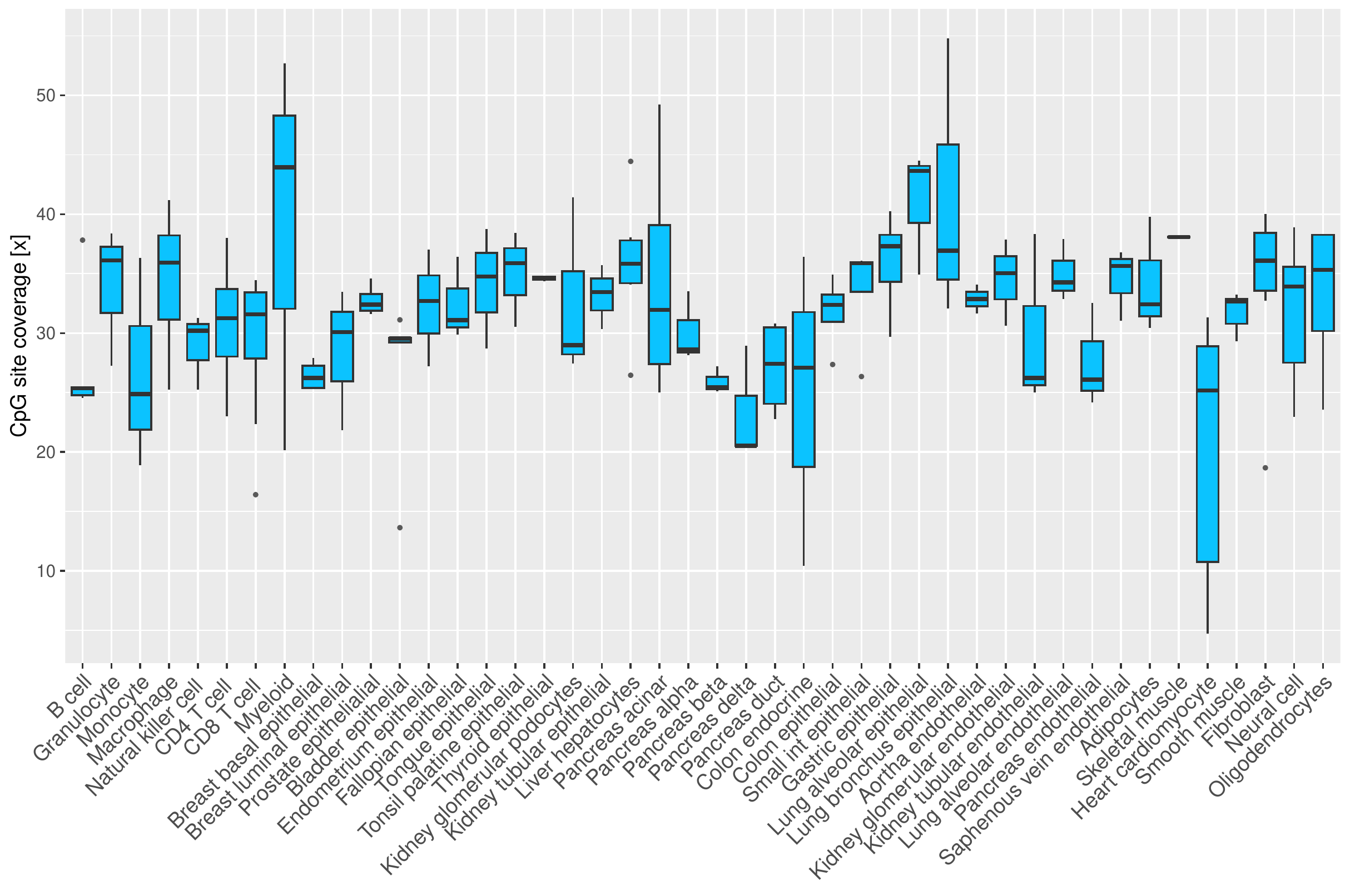}
    \caption{Sample CpG-site sequencing coverage across groups.}
    \label{fig:coverage}
\end{figure}


\begin{figure}[ht!]
    \hspace{2em}\includegraphics[width=1\textwidth]{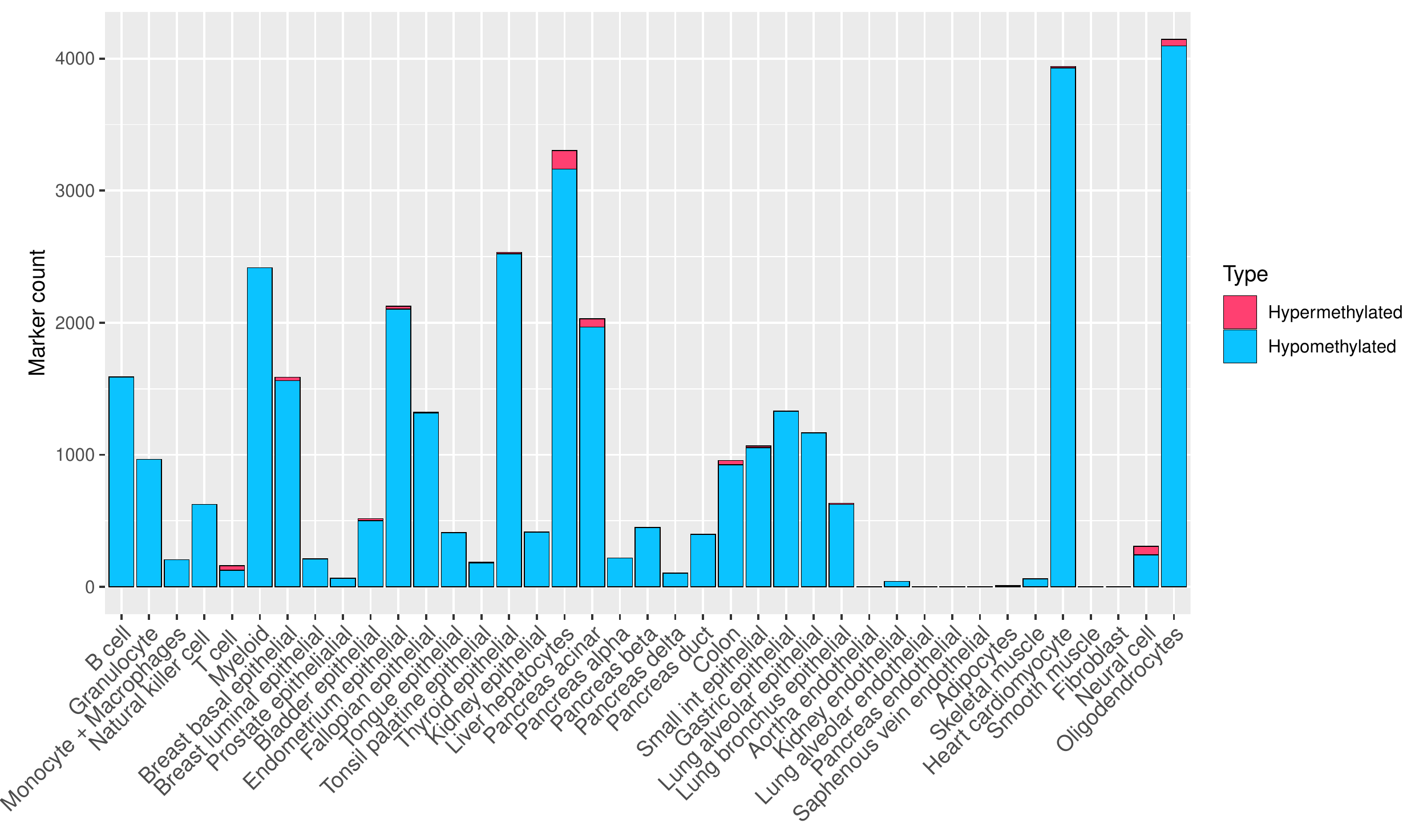}
    \caption{Number of hyper and hypomethylated UMRs when smooth muscle and fibroblast are separate groups.}
    \label{fig:marker_count_sep}
\end{figure}

\vspace{1em}
\begin{figure}[ht!]
    \hspace{-3.4em}
    \includegraphics[width=1.15\textwidth]{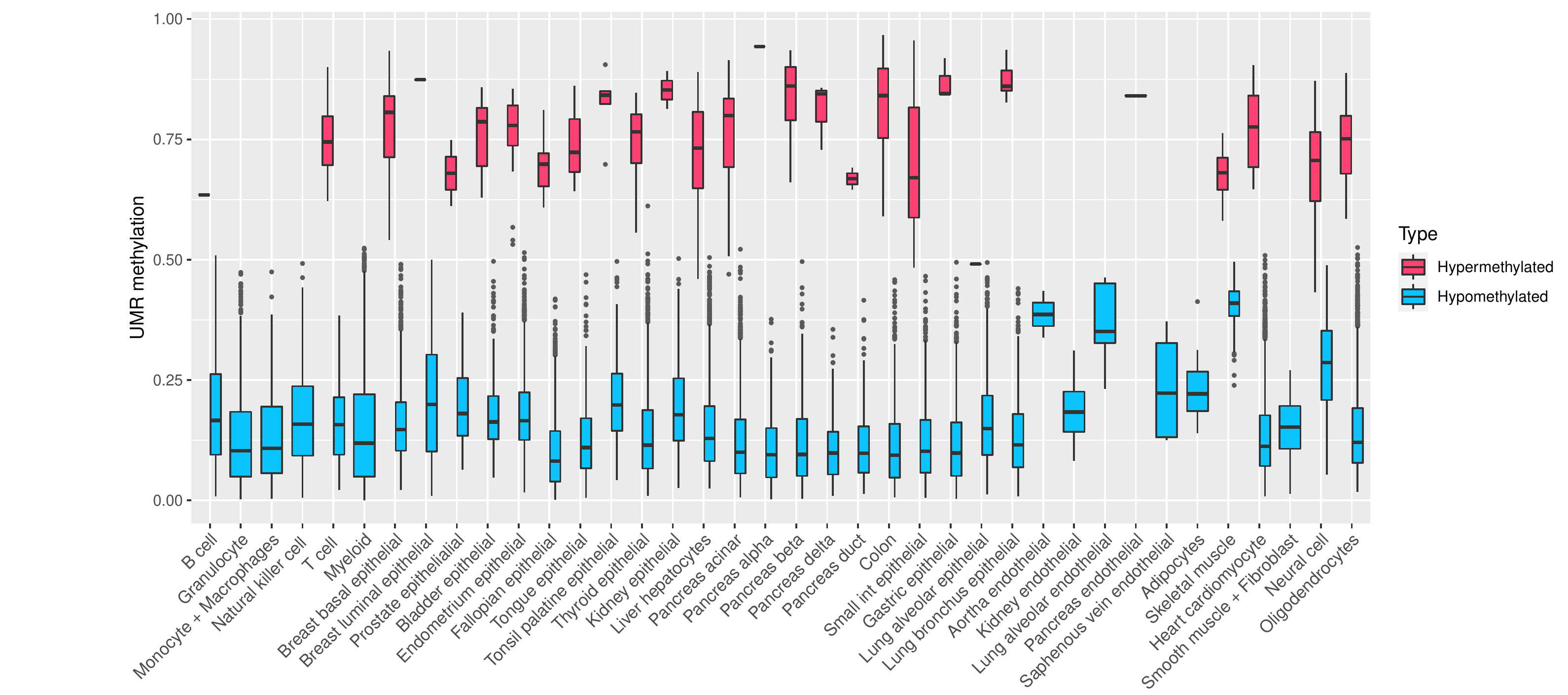}
    \caption{Methylation beta values of hyper and hypomethylated UMRs.}
    \label{fig:beta}
\end{figure}

\begin{figure}[ht!]
    \centering
    \hspace{4em}\includegraphics[width=0.8\textwidth]{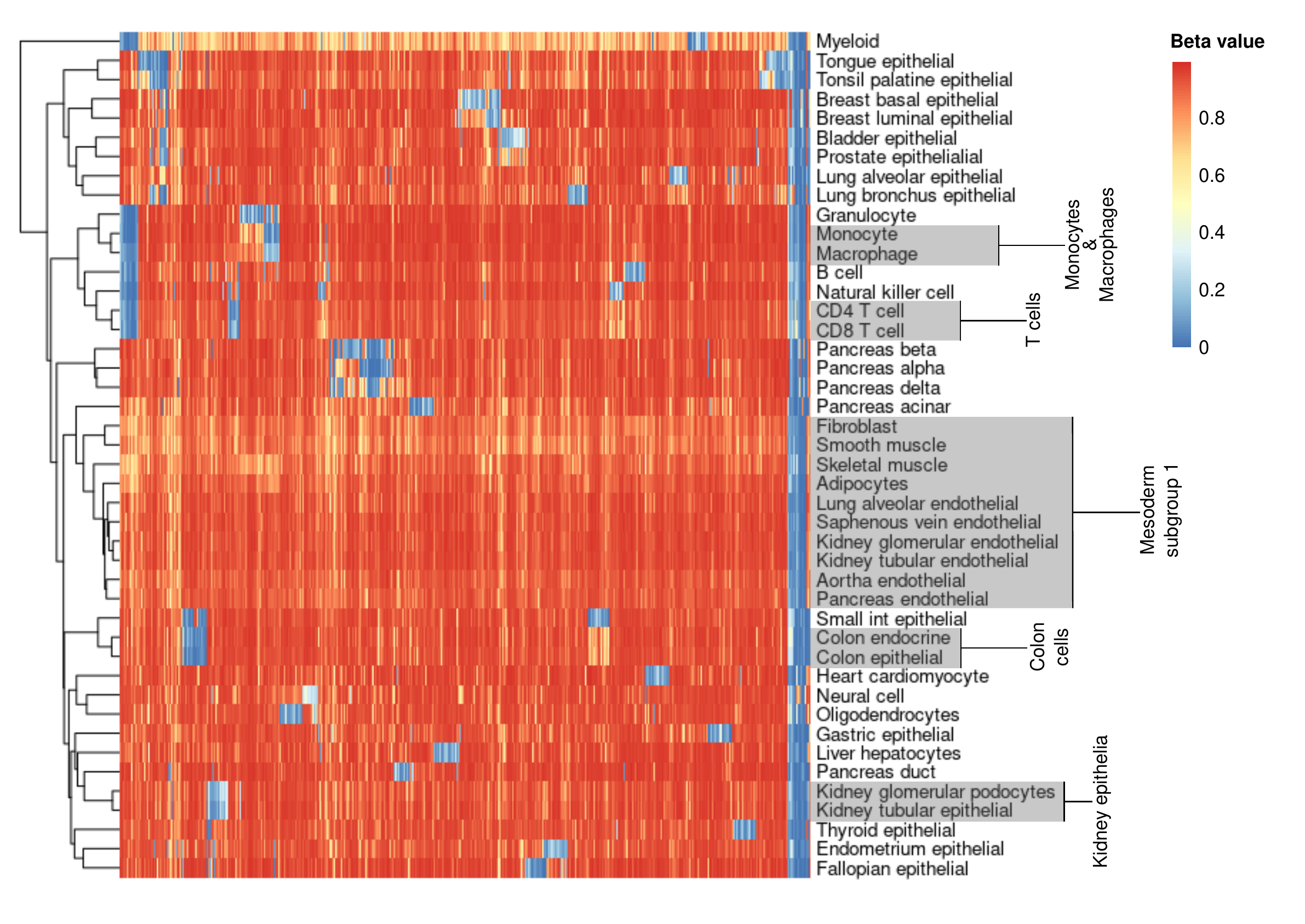}
    \caption{Beta values of all 10497 selected regions across all cell types. Clustering was performed using Manhattan distance and complete linkage. Joined cell types are shaded in gray.}
    \label{fig:all_basis}
\end{figure}

\begin{figure}[ht!]
    \hspace{2em}\includegraphics[width=1\textwidth]{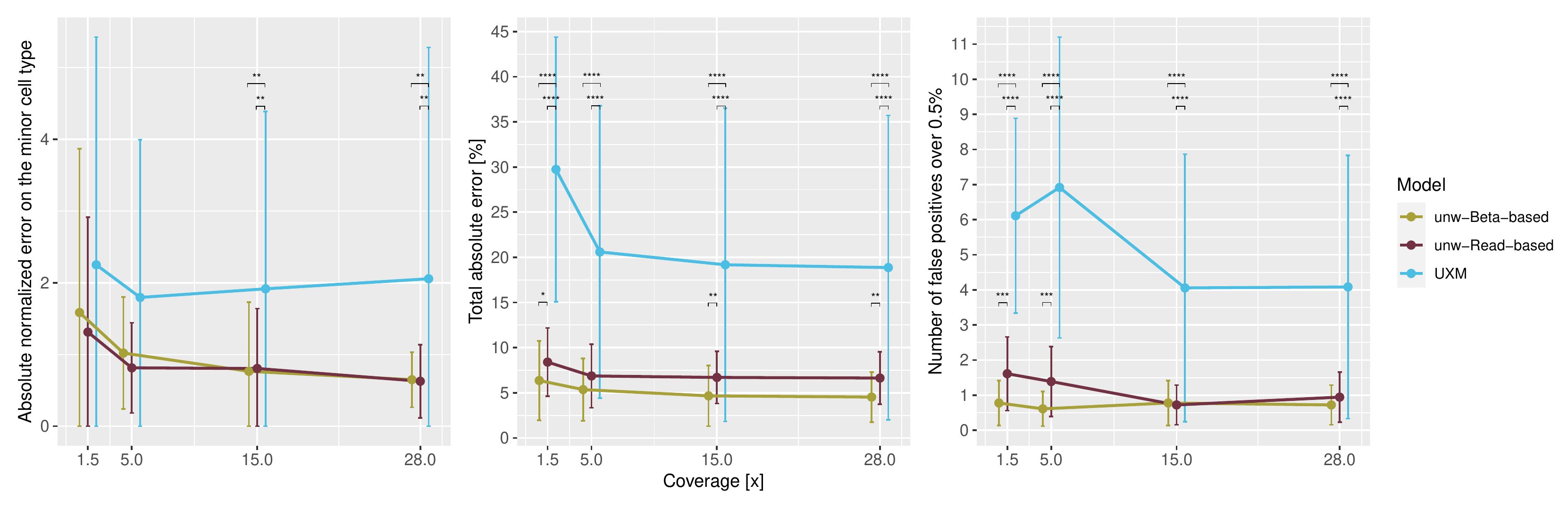}
    \caption{Aggregated deconvolution results when using non-weighted beta and read-based deconvolution. Mann-Whitney U test was used for statistical significance.}
    \label{fig:uxm_global}
\end{figure}

\begin{figure}[ht!]
    \hspace{2em}\includegraphics[width=1\textwidth]{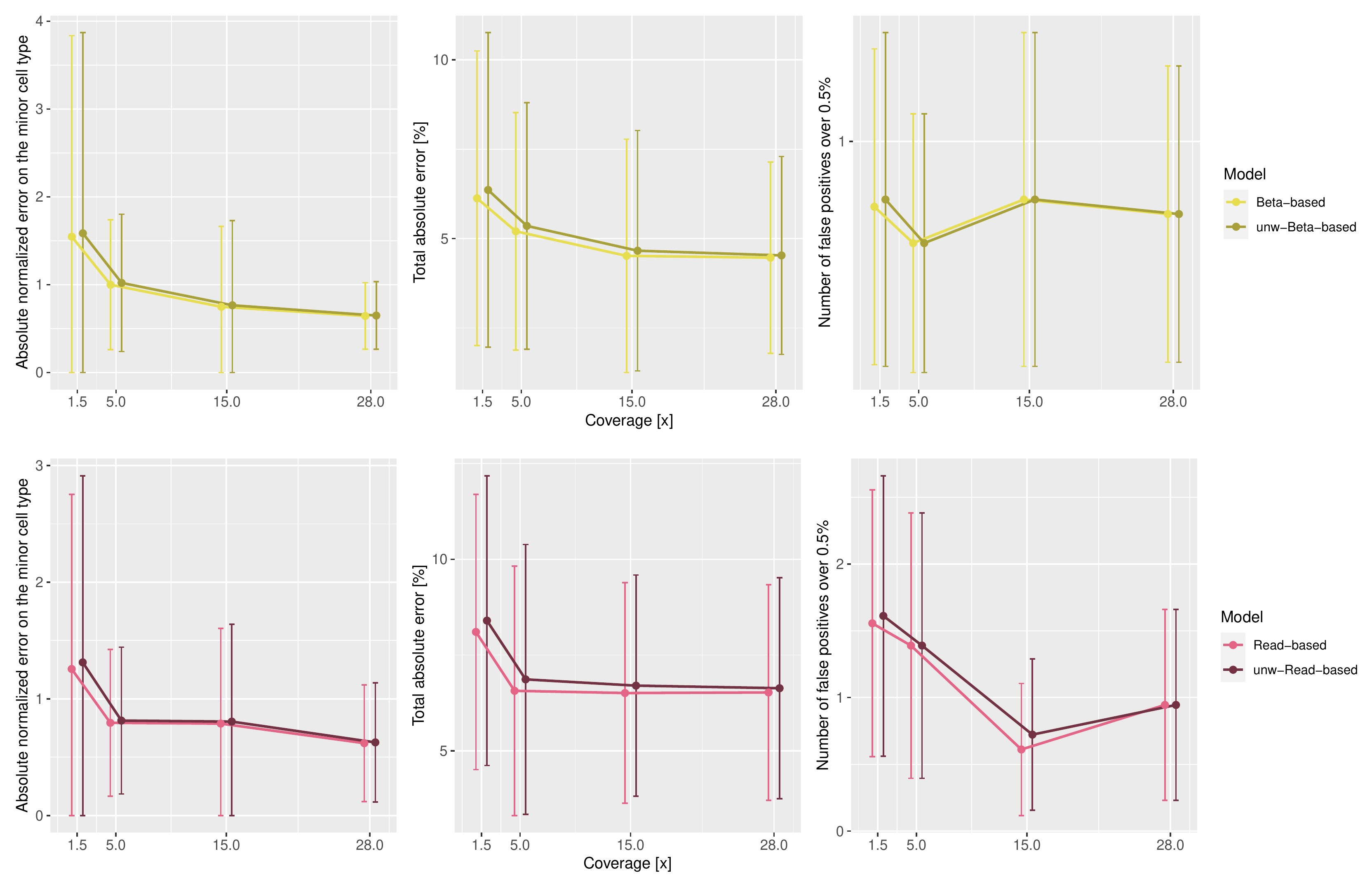}
    \caption{Aggregated deconvolution results using weighted and not weighted NNLS. No significant differences were identified.}
    \label{fig:joined_rnb}
\end{figure}

%% file: references.bib
@article{nature_intro_epi,
  title={Epigenetics and gene expression},
  author={Gibney, ER and Nolan, CM},
  journal={Heredity},
  year={2010},
  publisher={Nature Publishing Group}
}

@article{esteller2008epigenetics,
  title={Epigenetics in cancer},
  author={Esteller, Manel},
  journal={New England Journal of Medicine},
  year={2008},
  publisher={Mass Medical Soc}
}

@article{jang2017cpg,
  title={CpG and non-CpG methylation in epigenetic gene regulation and brain function},
  author={Jang, Hyun Sik and Shin, Woo Jung and Lee, Jeong Eon and Do, Jeong Tae},
  journal={Genes},
  year={2017},
  publisher={MDPI}
}

@article{laurent2010dynamic,
  title={Dynamic changes in the human methylome during differentiation},
  author={Laurent, Louise and Wong, Eleanor and Li, Guoliang and Huynh, Tien and Tsirigos, Aristotelis and Ong, Chin Thing and Low, Hwee Meng and Sung, Ken Wing Kin and Rigoutsos, Isidore and Loring, Jeanne and others},
  journal={Genome research},
  year={2010},
  publisher={Cold Spring Harbor Lab}
}

@article{illingworth2008novel,
  title={A novel CpG island set identifies tissue-specific methylation at developmental gene loci},
  author={Illingworth, Robert and Kerr, Alastair and DeSousa, Dina and J{\o}rgensen, Helle and Ellis, Peter and Stalker, Jim and Jackson, David and Clee, Chris and Plumb, Robert and Rogers, Jane and others},
  journal={PLoS biology},
  year={2008},
  publisher={Public Library of Science San Francisco, USA}
}

@article{strichman2002genome,
  title={A genome-wide screen for normally methylated human CpG islands that can identify novel imprinted genes},
  author={Strichman-Almashanu, Liora Z and Lee, Richard S and Onyango, Patrick O and Perlman, Elizabeth and Flam, Folke and Frieman, Matthew B and Feinberg, Andrew P},
  journal={Genome research},
  year={2002},
  publisher={Cold Spring Harbor Lab}
}

@article{cross1995cpg,
  title={CpG islands and genes},
  author={Cross, Sally H and Bird, Adrian P},
  journal={Current opinion in genetics \& development},
  year={1995},
  publisher={Elsevier}
}

@article{irizarry2009human,
  title={The human colon cancer methylome shows similar hypo-and hypermethylation at conserved tissue-specific CpG island shores},
  author={Irizarry, Rafael A and Ladd-Acosta, Christine and Wen, Bo and Wu, Zhijin and Montano, Carolina and Onyango, Patrick and Cui, Hengmi and Gabo, Kevin and Rongione, Michael and Webster, Maree and others},
  journal={Nature genetics},
  year={2009},
  publisher={Nature Publishing Group US New York}
}

@article{weber2007genomic,
  title={Genomic patterns of DNA methylation: targets and function of an epigenetic mark},
  author={Weber, Michael and Sch{\"u}beler, Dirk},
  journal={Current opinion in cell biology},
  year={2007},
  publisher={Elsevier}
}

@article{kulis2010dna,
  title={DNA methylation and cancer},
  author={Kulis, Marta and Esteller, Manel},
  journal={Advances in genetics},
  year={2010},
  publisher={Elsevier}
}

@article{wu2010active,
  title={Active DNA demethylation: many roads lead to Rome},
  author={Wu, Susan C and Zhang, Yi},
  journal={Nature reviews Molecular cell biology},
  year={2010},
  publisher={Nature Publishing Group UK London}
}

@article{guo2014active,
  title={Active and passive demethylation of male and female pronuclear DNA in the mammalian zygote},
  author={Guo, Fan and Li, Xianlong and Liang, Dan and Li, Tong and Zhu, Ping and Guo, Hongshan and Wu, Xinglong and Wen, Lu and Gu, Tian-Peng and Hu, Boqiang and others},
  journal={Cell stem cell},
  year={2014},
  publisher={Elsevier}
}

@article{onodera2021roles,
  title={Roles of TET and TDG in DNA demethylation in proliferating and non-proliferating immune cells},
  author={Onodera, Atsushi and Gonz{\'a}lez-Avalos, Edah{\'\i} and Lio, Chan-Wang Jerry and Georges, Romain O and Bellacosa, Alfonso and Nakayama, Toshinori and Rao, Anjana},
  journal={Genome Biology},
  year={2021},
  publisher={BioMed Central}
}

@article{lee2020key,
  title={The key role of DNA methylation and histone acetylation in epigenetics of atherosclerosis},
  author={Lee, Han-Teo and Oh, Sanghyeon and Yoo, Hyerin and Kwon, Yoo-Wook and others},
  journal={Journal of lipid and atherosclerosis},
  year={2020},
  publisher={The Korean Society of Lipid and Atherosclerosis}
}

@article{schaefer2010solving,
  title={Solving the Dnmt2 enigma},
  author={Schaefer, Matthias and Lyko, Frank},
  journal={Chromosoma},
  year={2010},
  publisher={Springer}
}

@article{chedin2002dna,
  title={The DNA methyltransferase-like protein DNMT3L stimulates de novo methylation by Dnmt3a},
  author={Chedin, Frederic and Lieber, Michael R and Hsieh, Chih-Lin},
  journal={Proceedings of the National Academy of Sciences},
  year={2002},
  publisher={National Acad Sciences}
}

@article{li2011distribution,
  title={Distribution of 5-hydroxy-methylcytosine in different human tissues},
  author={Li, Weiwei and Liu, Min},
  journal={Journal of nucleic acids},
  year={2011},
  publisher={Hindawi}
}

@article{peters2021calling,
  title={Calling differentially methylated regions from whole genome bisulphite sequencing with DMRcate},
  author={Peters, Timothy J and Buckley, Michael J and Chen, Yunshun and Smyth, Gordon K and Goodnow, Christopher C and Clark, Susan J},
  journal={Nucleic Acids Research},
  year={2021},
  publisher={Oxford University Press}
}

@article{benelli2021charting,
  title={Charting differentially methylated regions in cancer with Rocker-meth},
  author={Benelli, Matteo and Franceschini, Gian Marco and Magi, Alberto and Romagnoli, Dario and Biagioni, Chiara and Migliaccio, Ilenia and Malorni, Luca and Demichelis, Francesca},
  journal={Communications Biology},
  year={2021},
  publisher={Nature Publishing Group UK London}
}

@article{juhling2016metilene,
  title={Metilene: fast and sensitive calling of differentially methylated regions from bisulfite sequencing data},
  author={J{\"u}hling, Frank and Kretzmer, Helene and Bernhart, Stephan H and Otto, Christian and Stadler, Peter F and Hoffmann, Steve},
  journal={Genome research},
  year={2016},
  publisher={Cold Spring Harbor Lab}
}

@article{van2021targeted,
  title={A targeted solution for estimating the cell-type composition of bulk samples},
  author={van den Oord, Edwin JCG and Xie, Lin Y and Tran, Charles J and Zhao, Min and Aberg, Karolina A},
  journal={BMC bioinformatics},
  year={2021},
  publisher={BioMed Central}
}

@article{loyfer2023dna,
  title={A DNA methylation atlas of normal human cell types},
  author={Loyfer, Netanel and Magenheim, Judith and Peretz, Ayelet and Cann, Gordon and Bredno, Joerg and Klochendler, Agnes and Fox-Fisher, Ilana and Shabi-Porat, Sapir and Hecht, Merav and Pelet, Tsuria and others},
  journal={Nature},
  year={2023},
  publisher={Nature Publishing Group UK London}
}

@article{teschendorff2017comparison,
  title={A comparison of reference-based algorithms for correcting cell-type heterogeneity in Epigenome-Wide Association Studies},
  author={Teschendorff, Andrew E and Breeze, Charles E and Zheng, Shijie C and Beck, Stephan},
  journal={BMC bioinformatics},
  year={2017},
  publisher={BioMed Central}
}

@article{tang2018tumor,
  title={Tumor origin detection with tissue-specific miRNA and DNA methylation markers},
  author={Tang, Wei and Wan, Shixiang and Yang, Zhen and Teschendorff, Andrew E and Zou, Quan},
  journal={Bioinformatics},
  year={2018},
  publisher={Oxford University Press}
}

@article{chakravarthy2018pan,
  title={Pan-cancer deconvolution of tumour composition using DNA methylation},
  author={Chakravarthy, Ankur and Furness, Andrew and Joshi, Kroopa and Ghorani, Ehsan and Ford, Kirsty and Ward, Matthew J and King, Emma V and Lechner, Matt and Marafioti, Teresa and Quezada, Sergio A and others},
  journal={Nature communications},
  year={2018},
  publisher={Nature Publishing Group UK London}
}

@article{houseman2012dna,
  title={DNA methylation arrays as surrogate measures of cell mixture distribution},
  author={Houseman, Eugene Andres and Accomando, William P and Koestler, Devin C and Christensen, Brock C and Marsit, Carmen J and Nelson, Heather H and Wiencke, John K and Kelsey, Karl T},
  journal={BMC bioinformatics},
  year={2012},
  publisher={Springer}
}

@article{zhang2021emeth,
  title={EMeth: An EM algorithm for cell type decomposition based on DNA methylation data},
  author={Zhang, Hanyu and Cai, Ruoyi and Dai, James and Sun, Wei},
  journal={Scientific reports},
  year={2021},
  publisher={Springer}
}

@article{caggiano2021comprehensive,
  title={Comprehensive cell type decomposition of circulating cell-free DNA with CelFiE},
  author={Caggiano, Christa and Celona, Barbara and Garton, Fleur and Mefford, Joel and Black, Brian L and Henderson, Robert and Lomen-Hoerth, Catherine and Dahl, Andrew and Zaitlen, Noah},
  journal={Nature communications},
  year={2021},
  publisher={Nature Publishing Group UK London}
}

@article{levy2020methylnet,
  title={MethylNet: an automated and modular deep learning approach for DNA methylation analysis},
  author={Levy, Joshua J and Titus, Alexander J and Petersen, Curtis L and Chen, Youdinghuan and Salas, Lucas A and Christensen, Brock C},
  journal={BMC bioinformatics},
  year={2020},
  publisher={BioMed Central}
}

@article{arneson2020methylresolver,
  title={MethylResolver-a method for deconvoluting bulk DNA methylation profiles into known and unknown cell contents},
  author={Arneson, Douglas and Yang, Xia},
  journal={Communications biology},
  year={2020},
  publisher={Nature Publishing Group UK London}
}

@article{schmidt2020deconvolution,
  title={Deconvolution of cellular subsets in human tissue based on targeted DNA methylation analysis at individual CpG sites},
  author={Schmidt, Marco and Mai{\'e}, Tiago and Dahl, Edgar and Costa, Ivan G and Wagner, Wolfgang},
  journal={BMC biology},
  year={2020},
  publisher={BioMed Central}
}

@article{lutsik2017medecom,
  title={MeDeCom: discovery and quantification of latent components of heterogeneous methylomes},
  author={Lutsik, Pavlo and Slawski, Martin and Gasparoni, Gilles and Vedeneev, Nikita and Hein, Matthias and Walter, J{\"o}rn},
  journal={Genome biology},
  year={2017},
  publisher={Springer}
}

@article{fong2021determining,
  title={Determining subpopulation methylation profiles from bisulfite sequencing data of heterogeneous samples using DXM},
  author={Fong, Jerry and Gardner, Jacob R and Andrews, Jared M and Cashen, Amanda F and Payton, Jacqueline E and Weinberger, Kilian Q and Edwards, John R},
  journal={Nucleic acids research},
  year={2021},
  publisher={Oxford University Press}
}

@article{lee2019prism,
  title={PRISM: methylation pattern-based, reference-free inference of subclonal makeup},
  author={Lee, Dohoon and Lee, Sangseon and Kim, Sun},
  journal={Bioinformatics},
  year={2019},
  publisher={Oxford University Press}
}

@article{zheng2014methylpurify,
  title={MethylPurify: tumor purity deconvolution and differential methylation detection from single tumor DNA methylomes},
  author={Zheng, Xiaoqi and Zhao, Qian and Wu, Hua-Jun and Li, Wei and Wang, Haiyun and Meyer, Clifford A and Qin, Qian Alvin and Xu, Han and Zang, Chongzhi and Jiang, Peng and others},
  journal={Genome biology},
  year={2014},
  publisher={BioMed Central}
}

@article{sun2015base,
  title={Base resolution methylome profiling: considerations in platform selection, data preprocessing and analysis},
  author={Sun, Zhifu and Cunningham, Julie and Slager, Susan and Kocher, Jean-Pierre},
  journal={Epigenomics},
  year={2015},
  publisher={Future Medicine}
}

@article{jeong2022systematic,
  title={Systematic evaluation of cell-type deconvolution pipelines for sequencing-based bulk DNA methylomes},
  author={Jeong, Yunhee and de Andrade e Sousa, Lisa Barros and Thalmeier, Dominik and Toth, Reka and Ganslmeier, Marlene and Breuer, Kersten and Plass, Christoph and Lutsik, Pavlo},
  journal={Briefings in Bioinformatics},
  year={2022},
  publisher={Oxford University Press}
}

@article{keukeleire2022cell,
  title={Cell type deconvolution of methylated cell-free DNA at the resolution of individual reads},
  author={Keukeleire, Pia and Makrodimitris, Stavros and Reinders, Marcel},
  journal={bioRxiv},
  year={2022},
  publisher={Cold Spring Harbor Laboratory}
}

@article{scott2020identification,
  title={Identification of cell type-specific methylation signals in bulk whole genome bisulfite sequencing data},
  author={Scott, C Anthony and Duryea, Jack D and MacKay, Harry and Baker, Maria S and Laritsky, Eleonora and Gunasekara, Chathura J and Coarfa, Cristian and Waterland, Robert A},
  journal={Genome biology},
  year={2020},
  publisher={BioMed Central}
}

@article{thorvaldsdottir2013integrative,
  title={Integrative Genomics Viewer (IGV): high-performance genomics data visualization and exploration},
  author={Thorvaldsd{\'o}ttir, Helga and Robinson, James T and Mesirov, Jill P},
  journal={Briefings in bioinformatics},
  year={2013},
  publisher={Oxford University Press}
}

@article{cheng2015dna,
  title={DNA methylation and hydroxymethylation in stem cells},
  author={Cheng, Ying and Xie, Nina and Jin, Peng and Wang, Tao},
  journal={Cell biochemistry and function},
  year={2015},
  publisher={Wiley Online Library}
}

@article{quinlan2010bedtools,
  title={BEDTools: a flexible suite of utilities for comparing genomic features},
  author={Quinlan, Aaron R and Hall, Ira M},
  journal={Bioinformatics},
  year={2010},
  publisher={Oxford University Press}
}

@article{gao2020enhanceratlas,
  title={EnhancerAtlas 2.0: an updated resource with enhancer annotation in 586 tissue/cell types across nine species},
  author={Gao, Tianshun and Qian, Jiang},
  journal={Nucleic acids research},
  volume={48},
  number={D1},
  pages={D58--D64},
  year={2020},
  publisher={Oxford University Press}
}

@article{zeng2021silencerdb,
  title={SilencerDB: a comprehensive database of silencers},
  author={Zeng, Wanwen and Chen, Shengquan and Cui, Xuejian and Chen, Xiaoyang and Gao, Zijing and Jiang, Rui},
  journal={Nucleic acids research},
  year={2021},
  publisher={Oxford University Press}
}

@article{dor2018principles,
  title={Principles of DNA methylation and their implications for biology and medicine},
  author={Dor, Yuval and Cedar, Howard},
  journal={The Lancet},
  year={2018},
  publisher={Elsevier}
}

@article{kirillov1996role,
  title={A role for nuclear NF--$\kappa$B in B--cell--specific demethylation of the Ig$\kappa$ locus},
  author={Kirillov, Andrei and Kistler, Barbara and Mostoslavsky, Raul and Cedar, Howard and Wirth, Thomas and Bergman, Yehudit},
  journal={Nature genetics},
  year={1996},
  publisher={Nature Publishing Group}
}

@article{koukoura2012dna,
  title={DNA methylation in the human placenta and fetal growth},
  author={Koukoura, Ourania and Sifakis, Stavros and Spandidos, Demetrios A},
  journal={Molecular medicine reports},
  year={2012},
  publisher={Spandidos Publications}
}

@article{heinz2010simple,
  title={Simple combinations of lineage-determining transcription factors prime cis-regulatory elements required for macrophage and B cell identities},
  author={Heinz, Sven and Benner, Christopher and Spann, Nathanael and Bertolino, Eric and Lin, Yin C and Laslo, Peter and Cheng, Jason X and Murre, Cornelis and Singh, Harinder and Glass, Christopher K},
  journal={Molecular cell},
  year={2010},
  publisher={Elsevier}
}

@article{wang2012widespread,
  title={Widespread plasticity in CTCF occupancy linked to DNA methylation},
  author={Wang, Hao and Maurano, Matthew T and Qu, Hongzhu and Varley, Katherine E and Gertz, Jason and Pauli, Florencia and Lee, Kristen and Canfield, Theresa and Weaver, Molly and Sandstrom, Richard and others},
  journal={Genome research},
  year={2012},
  publisher={Cold Spring Harbor Lab}
}

@article{sun2022ctcf,
  title={CTCF and its partners: shaper of 3D genome during development},
  author={Sun, Xiaoyue and Zhang, Jing and Cao, Chunwei},
  journal={Genes},
  year={2022},
  publisher={MDPI}
}

@article{quaggin1999basic,
  title={The basic-helix-loop-helix protein pod1 is critically important for kidney and lung organogenesis},
  author={Quaggin, Susan E and Schwartz, Lois and Cui, Shiying and Igarashi, Peter and Deimling, Julie and Post, Martin and Rossant, Janet},
  journal={Development},
  year={1999},
  publisher={Company of Biologists The Company of Biologists, Bidder Building, 140 Cowley~…}
}

@article{roadmap2015integrative,
  title={Integrative analysis of 111 reference human epigenomes},
  author={Roadmap, Epigenomics C and Kundaje, A and Meuleman, W and Ernst, J and Bilenky, M and Yen, A and Heravi-Moussavi, A and Kheradpour, P and Zhang, Z and Wang, J and others},
  journal={Nature},
  year={2015}
}

@article{fernandez2016blueprint,
  title={The BLUEPRINT data analysis portal},
  author={Fernández, José María and de la Torre, Victor and Richardson, David and Royo, Romina and Puiggr{\`o}s, Montserrat and Moncunill, Valent{\'\i} and Fragkogianni, Stamatina and Clarke, Laura and Flicek, Paul and Rico, Daniel and others},
  journal={Cell systems},
  year={2016},
  publisher={Elsevier}
}

@article{zheng2012signification,
  title={Signification of hypermethylated in cancer 1 (HIC1) as tumor suppressor gene in tumor progression},
  author={Zheng, Jianghua and Xiong, Dan and Sun, Xueqing and Wang, Jinglong and Hao, Mingang and Ding, Tao and Xiao, Gang and Wang, Xiumin and Mao, Yan and Fu, Yuejie and others},
  journal={Cancer Microenvironment},
  year={2012},
  publisher={Springer}
}

@article{yao2019homeobox,
  title={The homeobox gene, HOXB13, regulates a mitotic protein-kinase interaction network in metastatic prostate cancers},
  author={Yao, Jiqiang and Chen, Yunyun and Nguyen, Duy T and Thompson, Zachary J and Eroshkin, Alexey M and Nerlakanti, Niveditha and Patel, Ami K and Agarwal, Neha and Teer, Jamie K and Dhillon, Jasreman and others},
  journal={Scientific reports},
  year={2019},
  publisher={Springer}
}

@article{gu2023rgreat,
  title={rGREAT: an R/bioconductor package for functional enrichment on genomic regions},
  author={Gu, Zuguang and H{\"u}bschmann, Daniel},
  journal={Bioinformatics},
  year={2023},
  publisher={Oxford University Press}
}

@article{mclean2010great,
  title={GREAT improves functional interpretation of cis-regulatory regions},
  author={McLean, Cory Y and Bristor, Dave and Hiller, Michael and Clarke, Shoa L and Schaar, Bruce T and Lowe, Craig B and Wenger, Aaron M and Bejerano, Gill},
  journal={Nature biotechnology},
  year={2010},
  publisher={Nature Publishing Group US New York}
}

@article{beltran2020circulating,
  title={Circulating tumor DNA profile recognizes transformation to castration-resistant neuroendocrine prostate cancer},
  author={Beltran, Himisha and Romanel, Alessandro and Conteduca, Vincenza and Casiraghi, Nicola and Sigouros, Michael and Franceschini, Gian Marco and Orlando, Francesco and Fedrizzi, Tarcisio and Ku, Sheng-Yu and Dann, Emma and others},
  journal={The Journal of clinical investigation},
  year={2020},
  publisher={Am Soc Clin Investig}
  }

@article{krueger2011bismark,
  title={Bismark: a flexible aligner and methylation caller for Bisulfite-Seq applications},
  author={Krueger, Felix and Andrews, Simon R},
  journal={bioinformatics},
  year={2011},
  publisher={Oxford University Press}
}

@article{danecek2021twelve,
  title={Twelve years of SAMtools and BCFtools},
  author={Danecek, Petr and Bonfield, James K and Liddle, Jennifer and Marshall, John and Ohan, Valeriu and Pollard, Martin O and Whitwham, Andrew and Keane, Thomas and McCarthy, Shane A and Davies, Robert M and others},
  journal={Gigascience},
  year={2021},
  publisher={Oxford University Press}
}

@article{yin2017impact,
  title={Impact of cytosine methylation on DNA binding specificities of human transcription factors},
  author={Yin, Yimeng and Morgunova, Ekaterina and Jolma, Arttu and Kaasinen, Eevi and Sahu, Biswajyoti and Khund-Sayeed, Syed and Das, Pratyush K and Kivioja, Teemu and Dave, Kashyap and Zhong, Fan and others},
  journal={Science},
  year={2017},
  publisher={American Association for the Advancement of Science}
}

@article{zabalza2015hoxb13,
  title={HOXB13 overexpression is an independent predictor of early PSA recurrence in prostate cancer treated by radical prostatectomy},
  author={Zabalza, Cristina Villares and Adam, Meike and Burdelski, Christoph and Wilczak, Waldemar and Wittmer, Corina and Kraft, Stefan and Krech, Till and Steurer, Stefan and Koop, Christina and Hube-Magg, Claudia and others},
  journal={Oncotarget},
  year={2015},
  publisher={Impact Journals, LLC}
}

@article{bird2002dna,
  title={DNA methylation patterns and epigenetic memory},
  author={Bird, Adrian},
  journal={Genes \& development},
  year={2002},
  publisher={Cold Spring Harbor Lab}
}

@article{okano1999dna,
  title={DNA methyltransferases Dnmt3a and Dnmt3b are essential for de novo methylation and mammalian development},
  author={Okano, Masaki and Bell, Daphne W and Haber, Daniel A and Li, En},
  journal={Cell},
  year={1999},
  publisher={Elsevier}
}

@article{debaun2003association,
  title={Association of in vitro fertilization with Beckwith-Wiedemann syndrome and epigenetic alterations of LIT1 and H19},
  author={DeBaun, Michael R and Niemitz, Emily L and Feinberg, Andrew P},
  journal={The American Journal of Human Genetics},
  year={2003},
  publisher={Elsevier}
}

@article{goldstone2004prader,
  title={Prader-Willi syndrome: advances in genetics, pathophysiology and treatment},
  author={Goldstone, Anthony P},
  journal={Trends in Endocrinology \& Metabolism},
  year={2004},
  publisher={Elsevier}
}

@article{temple2002transient,
  title={Transient neonatal diabetes, a disorder of imprinting},
  author={Temple, IK and Shield, JPH},
  journal={Journal of medical genetics},
  year={2002}
}

@article{liu2013epigenome,
  title={Epigenome-wide association data implicate DNA methylation as an intermediary of genetic risk in rheumatoid arthritis},
  author={Liu, Yun and Aryee, Martin J and Padyukov, Leonid and Fallin, M Daniele and Hesselberg, Espen and Runarsson, Arni and Reinius, Lovisa and Acevedo, Nathalie and Taub, Margaret and Ronninger, Marcus and others},
  journal={Nature biotechnology},
  year={2013},
  publisher={Nature Publishing Group US New York}
}

@article{ladd2014common,
  title={Common DNA methylation alterations in multiple brain regions in autism},
  author={Ladd-Acosta, Christine and Hansen, Kasper D and Briem, Eirikur and Fallin, M Daniele and Kaufmann, Walter E and Feinberg, Andrew P},
  journal={Molecular psychiatry},
  year={2014},
  publisher={Nature Publishing Group}
}

@article{de2014alzheimer,
  title={Alzheimer's disease: early alterations in brain DNA methylation at ANK1, BIN1, RHBDF2 and other loci},
  author={De Jager, Philip L and Srivastava, Gyan and Lunnon, Katie and Burgess, Jeremy and Schalkwyk, Leonard C and Yu, Lei and Eaton, Matthew L and Keenan, Brendan T and Ernst, Jason and McCabe, Cristin and others},
  journal={Nature neuroscience},
  year={2014},
  publisher={Nature Publishing Group US New York}
}

@article{esteller2007epigenetic,
  title={Epigenetic gene silencing in cancer: the DNA hypermethylome},
  author={Esteller, Manel},
  journal={Human molecular genetics},
  year={2007},
  publisher={Oxford University Press}
}

@article{chimonidou2013cst6,
  title={CST6 promoter methylation in circulating cell-free DNA of breast cancer patients},
  author={Chimonidou, Maria and Tzitzira, Alexandra and Strati, Areti and Sotiropoulou, Georgia and Sfikas, Costas and Malamos, Nikos and Georgoulias, Vasilis and Lianidou, Evi},
  journal={Clinical biochemistry},
  year={2013},
  publisher={Elsevier}
}

@article{peng2021circulating,
  title={Circulating tumor DNA and minimal residual disease (MRD) in solid tumors: current horizons and future perspectives},
  author={Peng, Yan and Mei, Wuxuan and Ma, Kaidong and Zeng, Changchun},
  journal={Frontiers in Oncology},
  year={2021},
  publisher={Frontiers Media SA}
}

@article{boer2021variations,
  title={Variations in DNA methylation and allograft rejection},
  author={Boer, Karin and Hesselink, Dennis A and Baan, Carla C},
  journal={Current Opinion in Organ Transplantation},
  year={2021},
  publisher={LWW}
}

@article{diaz2014liquid,
  title={Liquid biopsies: genotyping circulating tumor DNA.},
  author={Diaz, Jr LA and Bardelli, Alberto},
  journal={Journal of clinical oncology: official journal of the American Society of Clinical Oncology},
  year={2014},
  publisher={NIH Public Access}
}

@article{park2011promoter,
  title={Promoter CpG island hypermethylation during breast cancer progression},
  author={Park, So Yeon and Kwon, Hyeong Ju and Lee, Hee Eun and Ryu, Han Suk and Kim, Sung-Won and Kim, Jee Hyun and Kim, In Ah and Jung, Namhee and Cho, Nam-Yun and Kang, Gyeong Hoon},
  journal={Virchows Archiv},
  year={2011},
  publisher={Springer}
}

@article{issa2007dna,
  title={DNA methylation as a therapeutic target in cancer},
  author={Issa, Jean-Pierre J},
  journal={Clinical Cancer Research},
  year={2007},
  publisher={AACR}
}

@article{rauch2009human,
  title={A human B cell methylome at 100- base pair resolution},
  author={Rauch, Tibor A and Wu, Xiwei and Zhong, Xueyan and Riggs, Arthur D and Pfeifer, Gerd P},
  journal={Proceedings of the National Academy of Sciences},
  year={2009},
  publisher={National Acad Sciences}
}

@article{yu2014targeted,
  title={Targeted p16 Ink4a epimutation causes tumorigenesis and reduces survival in mice},
  author={Yu, Da-Hai and Waterland, Robert A and Zhang, Pumin and Schady, Deborah and Chen, Miao-Hsueh and Guan, Yongtao and Gadkari, Manasi and Shen, Lanlan and others},
  journal={The Journal of clinical investigation},
  year={2014},
  publisher={Am Soc Clin Investig}
}
